\documentclass[twocolumn]{aastex62}
\graphicspath{{./}{figures/}}
\usepackage[utf8]{inputenc}
\usepackage{graphicx}
\usepackage{dcolumn}
\usepackage{bm}
\usepackage{ulem} 
\usepackage{hyperref}

\usepackage{multirow,enumitem}
\usepackage{amsmath}
\usepackage{color}
\usepackage{xcolor}
\usepackage{multirow}

\definecolor{amethyst}{rgb}{0.6, 0.4, 0.8}
\definecolor{pink}{rgb}{0.8, 0.2, 0.8}

\newcommand{\be}{\begin{equation}}
\newcommand{\ee}{\end{equation}}

\newcommand{\siga}{\sigma_\alpha}
\newcommand{\sigb}{\sigma_\beta}
\newcommand{\E}[1]{\times 10^{#1}}
\newcommand{\lcdm}{$\Lambda$CDM}
\newcommand{\kmax}{k_{\mathrm{max}}}

\newcommand{\Cl}{C_\ell}

\newcommand{\Veff}{V_\textrm{eff}}
\newcommand{\Pobs}{P_\textrm{obs}}
\newcommand{\ds}{\displaystyle}
\newcommand{\hmpcinv}{\,h\,{\rm Mpc^{-1}}}
\newcommand{\Abary}{A_\mathrm{bary}}
\newcommand{\etao}{\eta_0}
\newcommand{\Mkz}{M_\textrm{nl}(k,z)}
\newcommand{\dPk}{\delta P(k,\mu,z)}
\newcommand{\NN}{\text{No Nuisance}}
\newcommand{\nn}{\text{No Nuisance}}
\begin{document}


\correspondingauthor{Xiaolei Li}
\email{lixiaolei@mail.bnu.edu.cn}

\title{The Quest for the Inflationary Spectral Runnings in the Presence of Systematic Errors}

\author{Xiaolei Li}
\affiliation{Department of Astronomy, Beijing Normal University, Beijing, 100875, China}
\affiliation{Department of Physics, University of Michigan, 450 Church St, Ann Arbor, MI 48109-1040, U.S.A.}
\author{Noah Weaverdyck}
\affiliation{Department of Physics, University of Michigan, 450 Church St, Ann Arbor, MI 48109-1040, U.S.A.}
\author{Saroj Adhikari}
\affiliation{Department of Physics, University of Michigan, 450 Church St, Ann Arbor, MI 48109-1040, U.S.A.}
\author{Dragan Huterer}
\affiliation{Department of Physics, University of Michigan, 450 Church St, Ann Arbor, MI 48109-1040, U.S.A.}
\author{Jessica Muir}
\affiliation{Department of Physics, University of Michigan, 450 Church St, Ann Arbor, MI 48109-1040, U.S.A.}
\author{Hao-Yi Wu}
\affiliation{Center for Cosmology and Astro-Particle Physics, The Ohio State University, Columbus, OH 43210-1117, U.S.A.}

\date{\today}
 
\begin{abstract}
Cosmological inflation predicts that the scalar spectral index ``runs'' with scale.  Constraints on the values of the spectral runnings, $\alpha_s\equiv \textrm{d} n_s/\textrm{d}\ln k$ and $\beta_s\equiv \textrm{d}\alpha_s/\textrm{d}\ln k$,  therefore provide a fundamental test of the physics of inflation. Here we study the feasibility of measuring the runnings when information from upcoming large-volume galaxy surveys is used to supplement the information provided by a CMB-S4 experiment, particularly focusing on the effect of including high-$k$, nonlinear scales. Since these measurements will be sensitive to modeling uncertainties for the nonlinear power spectrum, we  examine how three different ways of parameterizing those systematics---introducing zero, two, or several hundred nuisance parameters---affect constraints and protect against parameter biases. Considering statistical errors alone, we find that including strongly nonlinear scales can substantially tighten constraints.  However, these constraints weaken to levels not much better than those from a CMB-S4 experiment alone when we limit our analysis to scales where estimates are not strongly affected by systematic biases.  Given these considerations, near-future large-scale structure surveys are unlikely to add much information to the CMB-S4  measurement of the first running $\alpha_s$. There is more potential for improvement for the second running, $\beta_s$, for which  large-scale structure information will allow  constraints to be improved by a factor of 3--4 relative to using the CMB alone. Though these constraints are still above the value predicted by slow roll inflation, they do probe regions of parameter space relevant to nonstandard inflationary models with large runnings, for example those that can generate an appreciable abundance of primordial black holes.
\end{abstract}



\section{Introduction}\label{sec:intro}
Cosmological inflation~\citep{Guth:1980zm,Linde:1981mu,Albrecht:1982wi} has passed observational tests with flying colors: the combination of the cosmic microwave background (CMB) with measurements of large-scale structure (LSS) confirms that the geometry of the universe is nearly flat and that the spectrum of density fluctuations is almost scale-invariant~\citep{Bardeen:1983qw}. The super-horizon fluctuations observed in the temperature-polarization cross-correlation in the CMB behave in precisely the way that inflation predicts~\citep{Dodelson:2003ip}. Beyond these successes, the most important upcoming test of inflation is the search for the signature of primordial gravitational waves, which inflation generically predicts, in the CMB polarization. In this paper we study the prospects of another important test of inflation: the search for the running of the scalar spectral index.

The primordial power spectrum of curvature fluctuations can be parameterized by Taylor expanding about a pivot scale $k_*$ 
\be
\frac{k^3}{2 \pi^2}P_s(k) = A_s \left(\frac{k}{k_*}\right)^{(n_s-1) + \frac{1}{2} \alpha_s \ln (k/k_*)+ \frac{1}{6} \beta_s (\ln (k/k_*))^2 + ... },
\ee
where $A_s$ is the scalar amplitude, $n_s$ is the spectral index, and $\alpha_s$ and $\beta_s $ are its first and second derivatives, respectively, evaluated at the pivot scale $k_*$. Single-field slow-roll inflation models predict the power spectrum to be nearly scale invariant, i.e. $n_s \approx 1$, a prediction borne out through measurements of the CMB. The {\em Planck} experiment~\citep{Ade:2015lrj} has constrained these parameters for the $\Lambda$CDM$+\alpha_s$ model, measuring $n_s = 0.968 \pm 0.006$ and $\alpha_s = -0.003 \pm 0.007$ at the pivot $k_*=0.05$ Mpc$^{-1}$. Expanded to include the second running, the {\em Planck} constraints become $n_s = 0.959 \pm 0.006$, $\alpha_s = 0.009 \pm 0.010$, and $\beta_s = 0.025 \pm 0.013$ (see also ~\cite{Cabass:2016ldu}).

In single-field slow-roll inflationary models, the runnings are of the order $\alpha_s \sim (1-n_s)^2 \sim 10^{-3}$ and $\beta_s \sim (1-n_s)^3 \sim 4\times 10^{-5}$~\citep{Kosowsky:1995aa} (see also ~\cite{GarciaBellido:2014gna, Escudero:2015wba}), levels far below the sensitivity of the Planck satellite mission, but potentially reachable with new generations of CMB and LSS surveys. Detection of the runnings with magnitudes larger than these values would indicate that the mechanism that generated the primordial fluctuations cannot just be described by a single-field slow roll model~\citep{Easther:2006tv}. It is possible, for example, for large runnings to be generated by modulations to the inflationary potential \citep{Kobayashi:2010pz,Czerny:2014wua}. It has also been proposed that modulations resulting in a large value of $\beta_s \sim 10^{-3}$ could produce an appreciable number of primordial black holes (PBHs)~\citep{Drees:2011yz}; at $\beta_s \approx 0.03$, these PBHs would be large enough to be a dark matter candidate~\citep{Carr:2016drx,munoz2017towards}. Thus, even bounds on inflationary spectral runnings that are above the level needed to test single-field slow-roll inflation can provide valuable information.

\citet{munoz2017towards} investigated how well future surveys will be able to measure $\alpha_s$ and $\beta_s$, using a CMB Stage 4 (CMB-S4) experiment in combination with various LSS surveys. They find that even with the combination of a billion-object survey such as SKA, the runnings will only be measured to $\siga = 9.3\E{-4}$ and $\sigb = 2\E{-3}$, levels insufficient for a significant detection if the values are near those predicted by single-field slow-roll inflation (see ~\cite{Huang:2012mr,amendola2013cosmology,Basse:2014qqa} for other forecasts on spectral runnings constraints using CMB and future large-scale structure surveys). 
It is worth noting, however, that these forecasts only make use of LSS data that is comfortably in the linear regime ($k\lesssim 0.1\hmpcinv$). LSS surveys measure tracers of the matter power spectrum $P_m(k, z)$, and in principle can access information deep in the nonlinear regime, up to $k\simeq 1\hmpcinv$ and beyond. The combination of large scales accessed by the CMB and small scales accessed by the LSS is particularly important for constraining the spectral index and its running, as the long lever arm in wavenumber helps to break degeneracies with other cosmological parameters.

Using information from small scales (large $k$) introduces significant challenges, however. Fluctuations in matter density become large at small scales, so at some scale linear perturbation theory becomes insufficient to describe their evolution. There is a significant ongoing effort to improve our understanding of structure growth in this non-linear regime~\citep{Bernardeau:2001qr,smith2003stable,heitmann2009coyote,heitmann2010coyote,lawrence2010coyote}. Baryonic effects also become important at these scales and affect the power spectrum of large-scale structure tracers~\citep{Rudd:2007zx, vanDaalen:2011xb, Chisari:2018prw}. In addition, nonlinearities at small scales induce correlations between wavenumbers~\citep{Scoccimarro:1999kp}, so that the covariance of power spectra evaluated at two wavenumbers depends on the nontrivial matter trispectrum. 

It is therefore of fundamental importance to understand to what extent the small-scale systematics in the LSS can be parameterized and self-calibrated in order to utilize those scales in the search for $\alpha_s$ and $\beta_s$. 
The main goal of this paper is to assess how constraints on the runnings improve as LSS information at higher wavenumbers is added to the analysis. We investigate how the results are biased when the nonlinear regime is mismodeled, and how well this bias can be mitigated through the inclusion of nuisance parameters at small scales. 

The paper is organized as follows. In Sec. \ref{sec:methods}, we describe our methodology in detail: our fiducial cosmological model, the CMB and LSS surveys considered, and the Fisher matrix formalism we use for forecasting constraints. In Sec. \ref{sec:res}, we present and discuss our forecast for the spectral running $\alpha_s$ constraints using future galaxy surveys alone and in combination with CMB-S4. We then introduce the Fisher bias formalism for modeling systematic bias in cosmological parameters, 
and discuss the corresponding results for $\alpha_s$ in Sec. \ref{sec:sys}. In Sec. \ref{sec:beta}, we present our constraints and systematic bias results for the second spectral running, $\beta_s$. We summarize our findings and conclude in Sec. \ref{sec:conclusions}.

\section{Methods} \label{sec:methods}

In this section we describe our fiducial model for CMB and LSS observations and describe our forecasting methodology, which makes use of the Fisher matrix formalism to forecast the precision of measurements of the runnings. 

\begin{table}[!t]
\centering
\caption{Cosmological parameters, their fiducial values, and numerical derivative step sizes used for the Fisher matrix calculation. The last two parameters correspond to the Mead model for describing nonlinear effects.} \label{tab:paramsfid}
\begin{tabular}{ccc}
\hline
\hline\rule[-2mm]{0mm}{6mm} 
Parameter ($p_i$)  & Fiducial Value    & Step Size ($\Delta p_i$)     \\
\hline
\rule[-1mm]{0mm}{5mm} $\Omega_bh^2$  & 0.02222              & $\pm 1\%$      \\
\rule[-1mm]{0mm}{5mm}$\Omega_ch^2$  & 0.1197               & $\pm 1\%$      \\
\rule[-1mm]{0mm}{4mm}$\tau$         & 0.06                 & $\pm 1\%$      \\
\rule[-1mm]{0mm}{4mm}$H_0$          & 67.5                 & $\pm 1\%$         \\
\rule[-1mm]{0mm}{4mm}$n_s$          & 0.9655               & $\pm 1\%$      \\
\rule[-1mm]{0mm}{5mm}$10^{10}A_s$          & $21.96$      & $\pm 1\%$      \\
\rule[-1mm]{0mm}{4mm}$\alpha_s$     & 0                    & $\pm 1\times 10^{-3}$      \\
\rule[-1mm]{0mm}{5mm}$\beta_s$      & 0                    & $\pm 1\times 10^{-3}$      \\
\hline
\rule[-1mm]{0mm}{5mm}$\Abary$      & 3.13                    & $\pm 5\%$      \\
\rule[-1mm]{0mm}{5mm}$\etao$      & 0.6044                    & $\pm 5\%$      \\
\hline
\hline
\end{tabular}
\end{table}

\subsection{Fiducial model} \label{sec:model}
We assume a flat \lcdm{} cosmology with six parameters in addition to the spectral runnings: the physical baryon and CDM densities $\Omega_bh^2$ and $\Omega_ch^2$, the reionization optical depth $\tau$, the Hubble constant $H_0$, the scalar spectral index $n_s$, and the primordial power spectrum amplitude $A_s$. The values of these parameters in our fiducial model are listed in Table~\ref{tab:paramsfid}. 

\subsection{Modeling the CMB}

The CMB fluctuations have a wealth of information about the early universe, providing some of the tightest constraints for cosmology to date~\citep{Ade:2015xua}. 
The observed CMB angular power spectrum can be related to the primordial power spectrum $P_s(k)$ that sourced those fluctuations 
via
\be
\Cl^{XY} = \frac{2}{\pi}\int dk\,k^2 P_s(k) \Delta_\ell^X(k) \Delta_\ell^Y(k).
\ee
Labels $X$ and $Y$ can refer to temperature ($T$), polarization modes ($E, B$), or lensing potential ($d$), and $\Delta_\ell^X$ is the transfer function which encompasses both source and projection terms integrated over the line-of-sight.

Taking $T$ and $E$ as our observables, the observed angular power spectra can be represented as a vector $(\Cl^{TT},\, \Cl^{EE}, \,\Cl^{TE})$ with covariance matrix 
\begin{eqnarray} \label{eq:covCl}
&&\mathrm{Cov}_\ell = \ds\frac{2}{(2\ell+1)f_{sky}} \times \\[0.25cm]
&&\left(
\begin{array}{ccc}
(\tilde{C}_\ell^{TT})^2                ~~&  (\tilde{C}_\ell^{TE})^2               ~& \tilde{C}_\ell^{TT}\tilde{C}_\ell^{TE} \\[0.15cm]
(\tilde{C}_\ell^{TE})^2                ~~&  (\tilde{C}_\ell^{EE})^2               ~& \tilde{C}_\ell^{EE}\tilde{C}_\ell^{TE} \\[0.15cm]
\tilde{C}_\ell^{TT}\tilde{C}_\ell^{TE} ~~& \tilde{C}_\ell^{EE}\tilde{C}_\ell^{TE} ~& \frac{1}{2}[(\tilde{C}_\ell^{TE})^2+\tilde{C}_\ell^{TT}\tilde{C}_\ell^{EE}]
\end{array}\right)
\nonumber
\end{eqnarray}
where the auto power spectra include contributions from noise:

\begin{align}
\nonumber
\tilde{C}_\ell^{TT} &= C_\ell^{TT}+N_\ell^{TT} \\
\tilde{C}_\ell^{EE} &= C_\ell^{EE}+N_\ell^{EE} \\
\tilde{C}_\ell^{TE} &= C_\ell^{TE}\nonumber 
\end{align}

We adopt the same noise properties for a CMB-S4 experiment used by~\cite{munoz2017towards}:
\begin{equation}
{N_\ell^{TT}} = \Delta_T^2 
\exp\left [\ds\frac{\ell(\ell+1) \theta_{\text{FWHM}}^2}{8 \ln 2}\right  ]
\end{equation}
and
\begin{equation}
{N_\ell^{EE}} = 2\times N_\ell^{TT}  \ ,
\end{equation}
where the temperature sensitivity is $\Delta_T = 1 \mu$K-arcmin and the beam full-width-half-maximum is $\theta_{\rm{FWHM}} = 8.7\E{-4}$ radians~\citep{abazajian2016cmb}.
We assume a sky coverage $f_\mathrm{sky} = 0.4$ and  that the usable range of multipoles are $\ell \in [30, 3000]$ for $\Cl^{TT}$ and $\Cl^{TE}$, and  $\ell \in [30, 5000]$ for $\Cl^{EE}$. To represent additional constraints coming from low-$\ell$ polarization (e.g. from the Planck High Frequency Instrument) which break the degeneracy between $\tau$ and $A_s$~\citep{Aghanim:2016yuo}, we  include a Gaussian prior on $\tau$ with width $\sigma(\tau)=0.01$.

The nonlinearity of matter fluctuations affects the CMB power spectrum at small scales mainly through lensing. While the effect on the CMB lensing power spectrum from the large-scale structure bispectrum can be significant \citep{Bohm:2016gzt}, the corresponding changes in the TT, EE and TE angular power spectra are negligible \citep{Lewis:2016tuj}. We therefore do not consider the modeling uncertainties from nonlinear lensing effects on the CMB power spectra in this work.
\subsection{Modeling Large Scale Structure Surveys}
\label{sec:modeling}

LSS surveys utilize a variety of tracers in order to probe the growth of structure in the universe as a function of cosmic time, such as galaxies, quasars, and the Lyman-alpha forest. These measurements, in turn, enable strong constraints to be placed on both early- and late-universe parameters
~\citep{tegmark2004cosmological,samushia2012interpreting,alam2017clustering}.

In the linear regime, the matter power spectrum can be computed for a given cosmology using Boltzmann codes such as \texttt{CAMB}~\citep{lewis538efficient} or \texttt{CLASS}~\citep{blas2011cosmic}. On smaller scales where linear perturbation theory breaks down, one must resort to other methods. These may include N-body or hydrodynamical simulations, or else  semi-analytic prescriptions, for example ones based on the halo model of LSS~\citep{Seljak:2000gq,Peacock:2000qk,Cooray:2002dia}. 
However these methods are not guaranteed to capture all the relevant physics. 
The presence of redshift space distortions (RSD), which render the power spectrum observed in redshift space anisotropic, further complicates matters.

Because we aim to investigate the impact of systematic errors on constraints from LSS, and those are mainly due to modeling uncertainties at small scales, we parameterize the observed galaxy power spectrum in a way that allows us to generically encapsulate modifications to our fiducial power spectrum due to nonlinear effects. Following~\citet{seo2007improved}, we write the redshift-space power spectrum of tracer $X$ as 
\begin{equation}
\begin{aligned}
\label{eq:Pobs}
P_{\rm obs}(k,\mu, z) &=
 b_X^2(z)\left (1+\frac{f(z)}{b_X(k,z)}\mu^2\right )^2\\[0.2cm]
&\times P_m(k,z){\,\exp\left[- \frac{k^2\mu^2\sigma_{v}^2}{H_0^2}  \right]}\,\Mkz
\end{aligned}
\end{equation}
where $P_m(k,z)$ is the matter power spectrum from \texttt{CAMB} with nonlinear corrections from  \texttt{HMcode}~\citep{mead2015accurate}, $\mu$ is the cosine of the angle between the line connecting galaxy pairs and the line of sight and $f(z)=\textrm{d}\ln D/\textrm{d}\ln a$ is the logarithmic derivative of the linear growth factor. The exponential term, featuring the velocity dispersion $\sigma_v$, models the power suppression along the line-of-sight at small scales due to redshift-space distortions (the so-called Figures-of-God effect).
Here $\sigma_v$ is calculated using the virial scaling relation from \cite{Evrard08}, evaluated at the  characteristic mass of collapsed halos ($M_*$).
We find that the effect only has a minor impact, slightly increasing the forecasted errors at  $\kmax>1 \hmpcinv$.
The impact of baryons and other effects on nonlinear scales (henceforth \textit{nonlinear effects}) are accounted for by the as-yet undefined function $M_\textrm{nl}(k,z)$. The term $b_X^2$ describes the linear galaxy bias for galaxy population $X$, which we define to have the redshift dependence of $b^2(z) = b_0^2(1+z)$. We marginalize over the amplitude $b_0$ when determining cosmological parameter constraints, and absorb any scale-dependent bias effects into $\Mkz$.

We now turn to the ``nonlinear'' function $\Mkz$. We consider three treatments, in order of increasing complexity:
\begin{enumerate}[leftmargin=*]
\item {\bf {$\NN$} model:} The simplest case is the trivial one where the nonlinear power is assumed to be modeled perfectly by the modified halo model prescription in {\tt HMCode} and there is no scale-dependent bias. This corresponds to $\Mkz = 1$, with no additional nuisance parameters. We refer to this as the \textit{$\NN$} model. 
\item {\bf Mead model:} The next model for $\Mkz$ is the one presented by~\citet{mead2015accurate}, in which the modifications to nonlinear power due to baryonic feedback effects are parametrized using two parameters [$A_b$ and $\eta_0$] (\textit{Mead parameters}). In this case, 
\begin{eqnarray}
\Mkz &=& \frac{P_\textrm{Mead}(k, z, A_b, \eta_0)}{P_\textrm{Mead, DMonly}(k, z)}
\label{eq:Mead}
\end{eqnarray}
where ``DMonly'' refers to the default Mead parameter values of $A_b=3.13$ and $\eta_0=0.6044$.
\item {\bf Many Free Parameter (MFP) model:} The final model for the nonlinearities is a much more agnostic prescription similar to~\citet{bielefeld2015cosmological}, in which $\Mkz$ is allowed to float freely in bins of wavenumber $k$ and smoothly, as a low-order power-law, in redshift. 
Since at low $k$ the power spectrum is well determined theoretically, we allow $\Mkz$ to vary only for $k$ at the quasi-linear regime and above, setting it to unity at large scales. 

We therefore have
\begin{eqnarray}
\label{eq:MFP}
  \Mkz &=&
  \begin{cases}
   		(1+c_{1,k}z+c_{2,k}z^2)B_k & \text{if $k> 0.1$} \\
   \quad \quad \quad 1 & \text{if $k\leq0.1$} 
  \end{cases}
\end{eqnarray}
where $k$ has units $\hmpcinv$, and $B_k$, $c_{1,k}$ and $c_{2,k}$ are free parameters. One set of \{$B_k$, $c_{1,k}$, $c_{2,k}$\} is specified in each wavenumber bin of width $\Delta \ln k=0.05\hmpcinv$. This bin width is fixed, so as the maximum wavenumber $\kmax$ is raised, the number of $k$ bins increases, and consequently so does the number of nuisance parameters. The total number of nuisance parameters in $\Mkz$ thus ranges from 0 to 279 as $\kmax$ is varied from $0.1$ to $10\hmpcinv$, and hence we refer to this as the \textit{Many Free Parameter} (\textit{MFP}) model.
\end{enumerate}

The covariance between the observed power spectrum at wave numbers $k_{\alpha}$ and $k_{\beta}$ is given as the sum of the ``unconnected" part, which is diagonal in the two wavenumbers, and the connected contribution given by the full trispectrum:
\begin{equation}\label{eq:Cov}
\left[{\rm{Cov}}\right]_{k_\alpha,k_\beta} = \frac{8\pi^2 [\Pobs(k_\alpha,\mu,z)]^2}{\Veff(k_\alpha,\mu,z)k_\alpha^2 \Delta k_\alpha}\delta_{k_\alpha,k_\beta}+T_{k_\alpha, k_\beta}.
\end{equation}
The effective volume of the survey varies with redshift and is given by
\begin{equation}
\Veff(k,\mu,z) = V(z)_\textrm{survey}\left[\frac{n(z)\Pobs(k,\mu,z)}{1+n(z)\Pobs(k,\mu,z)}\right]^2
\end{equation}
where $n(z)$ is the galaxy number density of each redshift bin and $V(z)_\textrm{survey}$ is the volume in $[h^{-1} {\rm{Mpc}}]^3$,
\begin{equation}
V(z)_\textrm{survey} = \int_{z_{\rm min}}^{z_{\rm max}}\Omega_\textrm{survey} \frac{r(z')^2}{H(z')} dz'.
\end{equation}
Here, $r(z)$ is the comoving distance, $H(z)$ is the Hubble parameter, and $\Omega_\textrm{survey}$ is the sky coverage of the survey in steradians. The term  $T_{k_\alpha,k_\beta}$ is the contribution from the trispectrum due to the non-Gaussian nature of the matter field,
\begin{equation}
{T_{k_\alpha,k_\beta}} = \int_{k_\alpha} \frac{d^3\vec{k_1}}{4 \pi k_\alpha^2 \Delta k_\alpha} \int_{k_\beta} \frac{d^3\vec{k_2}}{{4 \pi k_\beta^2 \Delta k_\beta}}T(\vec{k_1},-\vec{k_1},\vec{k_2},-\vec{k_2}).
\end{equation}
We obtain $T_{k_\alpha,k_\beta}$ with the same calculation method described by~\citet{wu2013impact}, who use the halo model to calculate the trispectrum, showing that it is dominated by the one-halo term. We refer the interested reader to that work for details. 

In their spectral running constraint forecasts, ~\citet{munoz2017towards} consider a wide survey like the Dark Energy Spectroscopic Instrument (DESI)~\citep{aghamousa2016desi} as well as a deep and narrow survey similar to the Wide Field Infrared Survey Telescope (WFIRST)~\citep{Spergel:2015sza}, finding that they improve constraints on the runnings by $\sim20\%$ and $30\%$, respectively, when added to data from a CMB-S4 experiment. Here we take a Euclid-like survey to be our fiducial survey, and we include a DESI-like survey for comparison. 

\begin{figure}
\centering
\includegraphics[width=0.950\linewidth]{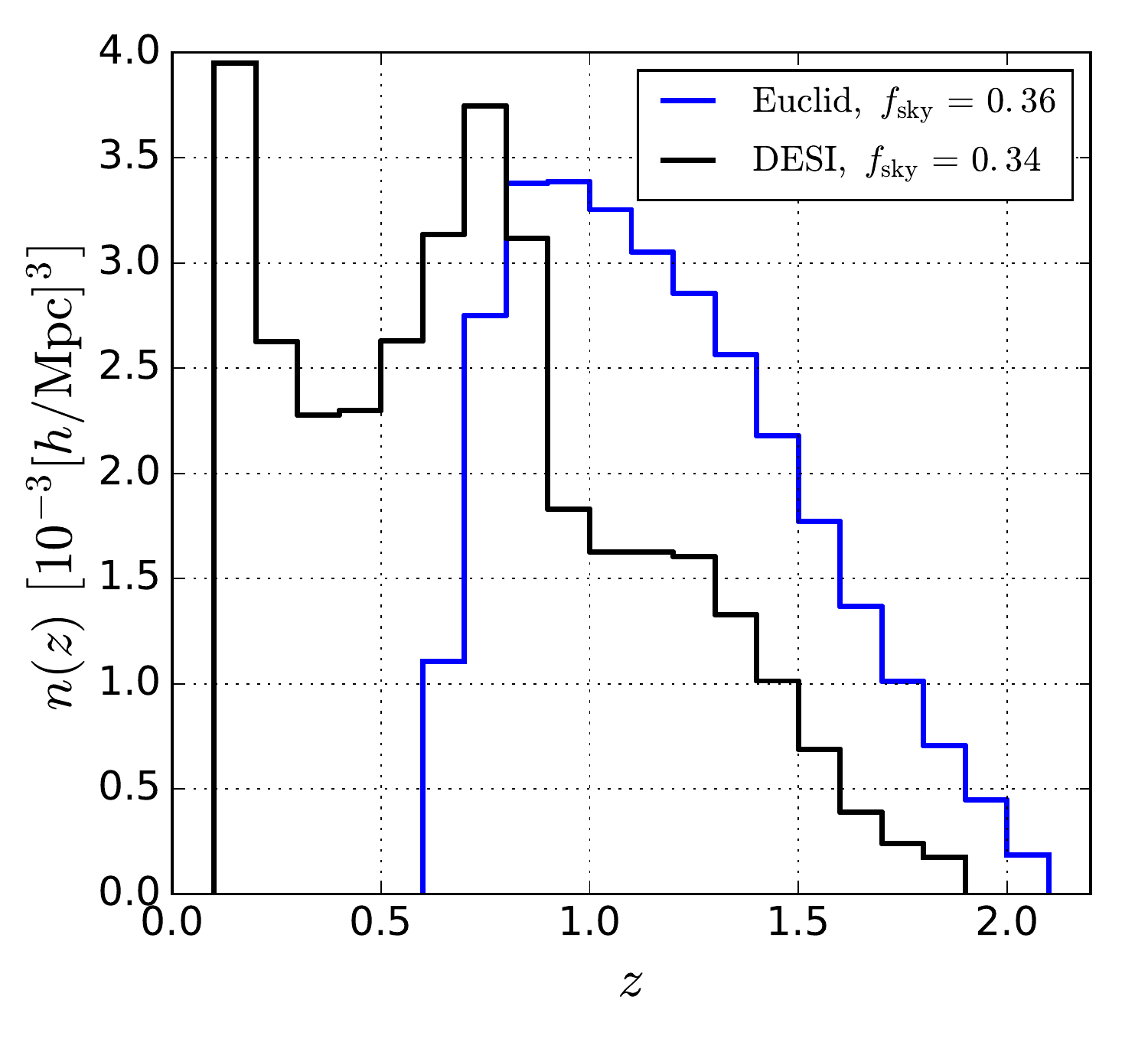}
\caption{Galaxy number density $n(z)$ of Euclid and DESI in each redshift bin. The features in the DESI $n(z)$ are due to the fact that the sample is a combination of several populations of sources.}
\label{fig:dndz}
\end{figure}

\bigskip
\textbf{Euclid:}
Euclid~\citep{laureijs2011euclid} is a proposed space-based LSS survey with large sky coverage and a deep redshift distribution, which should provide excellent constraints on the evolution of dark energy.
We use the spectroscopic sample defined in~\cite{laureijs2011euclid}, assuming $15000$ deg$^2$ ($f_\textrm{sky}\approx 0.36$) and a total of $50$ million galaxies. We use the redshift bins given in Table~VI of~\cite{font2014desi}, with thickness $\Delta z = 0.1$ in the range $z \in [0.6, 2.1]$. We infer the effective number density in each bin as $n(z) =\bar{n}P_{0.14,0.6}(z) / P_\mathrm{obs}(k=0.14h \mathrm{Mpc}^{-1}, \mu=0.6, z)$, where $\bar{n}P_{0.14,0.6}(z)$ is a quantity reported by~\cite{font2014desi} and $P_\mathrm{obs}$ is calculated via Eq.~(\ref{eq:Pobs}). The resulting $n(z)$ is shown in Fig.~\ref{fig:dndz}.

\bigskip
\textbf{DESI:}
The Dark Energy Spectroscopic Instrument (DESI~\citep{aghamousa2016desi}) is a Stage-IV ground-based spectroscopy experiment at Mayall telescope in Arizona, which will target multiple tracer populations over $14,000$ deg$^2$ ($f_\textrm{sky}\approx 0.34$) with good signal to noise out to $z\lesssim 1.5$. Here too we adopt the distribution given in~\cite{font2014desi}, which combines projections for the populations of Luminous Red Galaxies (LRGs), Emission Line Galaxies (ELGs) and quasars (QSOs) into estimates of $\bar{n}P_{0.14,0.6}(z)$ in redshift bins of $\Delta z = 0.1$ in the range $z \in [0.1, 1.9]$. We calculate an effective $n(z)$ for each bin in the same way as with the Euclid-like projections, and show them in Fig.~\ref{fig:dndz}. We assume that the Euclid-like and DESI-like experiments do not overlap and we combine their information by summing the Fisher matrices, as we describe below.

\subsection{Forecasting} \label{sec:fisher}
We forecast uncertainties of cosmological parameters as a function of $\kmax$ using a Fisher matrix analysis.
The Fisher matrix formalism is an extremely simple and efficient method to estimate the errors on model parameters given a set of data
\citep{Albrecht:2009ct,tegmark1997fisher}. If one approximates the likelihood as a multi-variate Gaussian in the parameters around its peak, the resulting Hessian (matrix of second derivatives) can be used to calculate the forecasted uncertainties in the cosmological parameters. The better the actual constraints on the parameters are, the closer the likelihood function is to a Gaussian distribution, and the more accurate the Fisher matrix approximation is.  To the extent that we are assuming powerful future surveys with small errors on most parameters, the Fisher matrix approximation should be excellent. More importantly, given that our MFP systematics case contains up to $\sim$$300$ parameters, a Fisher forecast is the \textit{only} reasonably straightforward way to estimate the errors.

Under the assumption of Gaussian perturbations and Gaussian noise, the Fisher Matrix for CMB temperature and polarization anisotropies~\citep{seljak1997measuring,zaldarriaga1997all,eisenstein1999cosmic} can be written  as
\begin{equation}
F_{ij}^{\rm{CMB}} = \sum_\ell {A_{\ell,i}(\mathrm{Cov}_\ell)^{-1}[A_{\ell,{j}}]^T},
\end{equation}
where
\begin{eqnarray}
A_{\ell,i}& = & \left(
\begin{array}{ccc}
\ds\frac{\partial C_\ell^{TT}}{\partial p_i}, & 
\ds\frac{\partial C_\ell^{EE}}{\partial p_i}, & 
\ds\frac{\partial C_\ell^{TE}}{\partial p_i} 
\end{array}\right)
\end{eqnarray}
and the covariance is given by Eq.~(\ref{eq:covCl}).
The Fisher matrix for the observed LSS power spectrum is
\begin{equation}\label{eq:Fisher}
\begin{aligned}
F_{ij}^{\rm{LSS}}\,&= \,\sum_z\sum_\mu d\mu\sum_{{k_{\alpha}},k_{\beta}}
\frac{\partial P(k_{\alpha},\mu,z)}{\partial p_i}\\[0.20cm]
&\times \left[\rm{Cov}^{-1}\right]_{k_\alpha,k_\beta}\frac{\partial P(k_{\beta},\mu,z)}{\partial p_j},
\end{aligned}
\end{equation}
where the sums are over all bins in $z$, $\mu$, and $k$, 
and $p_i$ runs over the cosmological parameters $\{\Omega_bh^2, \Omega_ch^2, \tau, H_0, n_s, A_s, \alpha_s,\beta_s\}$ as well as the linear bias parameter $b_0$ and the nuisance parameters in every $k$-bin, $\{B_k, c_{1,k}, c_{2,k}\}$.
We define $k$ bins logarithmically, with $\Delta \ln k = 0.05$ 
in the range $\kmax \in [0.1, 10] \hmpcinv$, and bin $\mu$ in 11 evenly spaced bins from $-1$ to $1$.
 
Forecasts for a combination of experiments can be calculated by summing their Fisher matrices, and a forecast for the lower bound on the the error for a given parameter is given by the Cramer-Rao inequality
\begin{displaymath}
\sigma(p_i) \geq  \left\{ \begin{array}{ll}
{\sqrt{{(F^{-1}})_{ii}}}  & \textrm{(marginalized)}\\[0.2cm]
 1/{\sqrt{F_{ii}}} & \textrm{(unmarginalized)}.
 \end{array} \right.
\end{displaymath}

\section{Results} \label{sec:res}

We now present the principal results. To give an idea of the  approximate overall level of constraint on the cosmological parameters, we summarize the fiducial constraints for our CMB-S4 forecast on the spectral runnings: when fixing $\beta_s=0$, we obtain marginalized error on the spectral running of $\siga=3.0\times 10^{-3}$. When allowing $\beta_s$ to vary, we find $\siga=3.4\times 10^{-3}, \sigb=8.0\times 10^{-3}$. (All constraints listed are the marginalized error, unless otherwise noted.) These constraints are similar to those of~\citet{munoz2017towards}, although slightly weaker because we do not include lensing information.

We now turn to the main goal of the paper: exploring whether and how adding information from LSS improves constraints on the spectral runnings. We first consider galaxy clustering alone, and then in conjunction with CMB-S4.

\begin{figure}
\includegraphics[width=\linewidth]{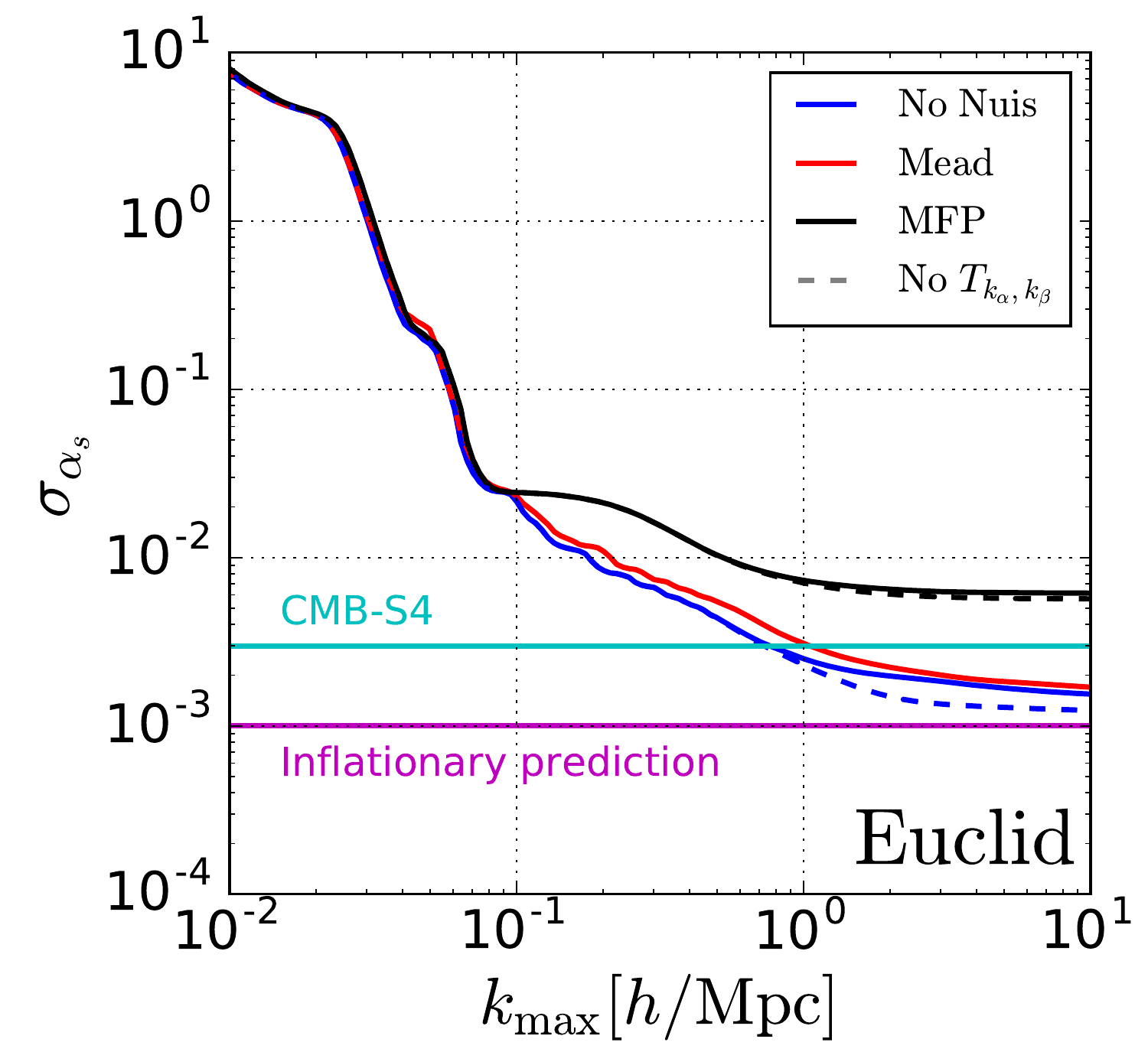}
\caption{Error ($1\sigma$ here and everywhere) in the spectral running $\alpha_s$ as a function of $\kmax$, evaluated for our fiducial Euclid-like survey, assuming $\beta_s = 0$. The legend shows our assumption about modeling of the systematics, while 
$T_{k_\alpha, k_\beta}$ refers to the inclusion of the trispectrum to the data covariance. Note that here and in subsequent plots, the value of the running denoted as the ``Inflationary prediction" (purple horizontal line) is only approximate.}
\label{fig:error_Euclid}
\end{figure}

\begin{figure*}
\includegraphics[width=0.85\linewidth]{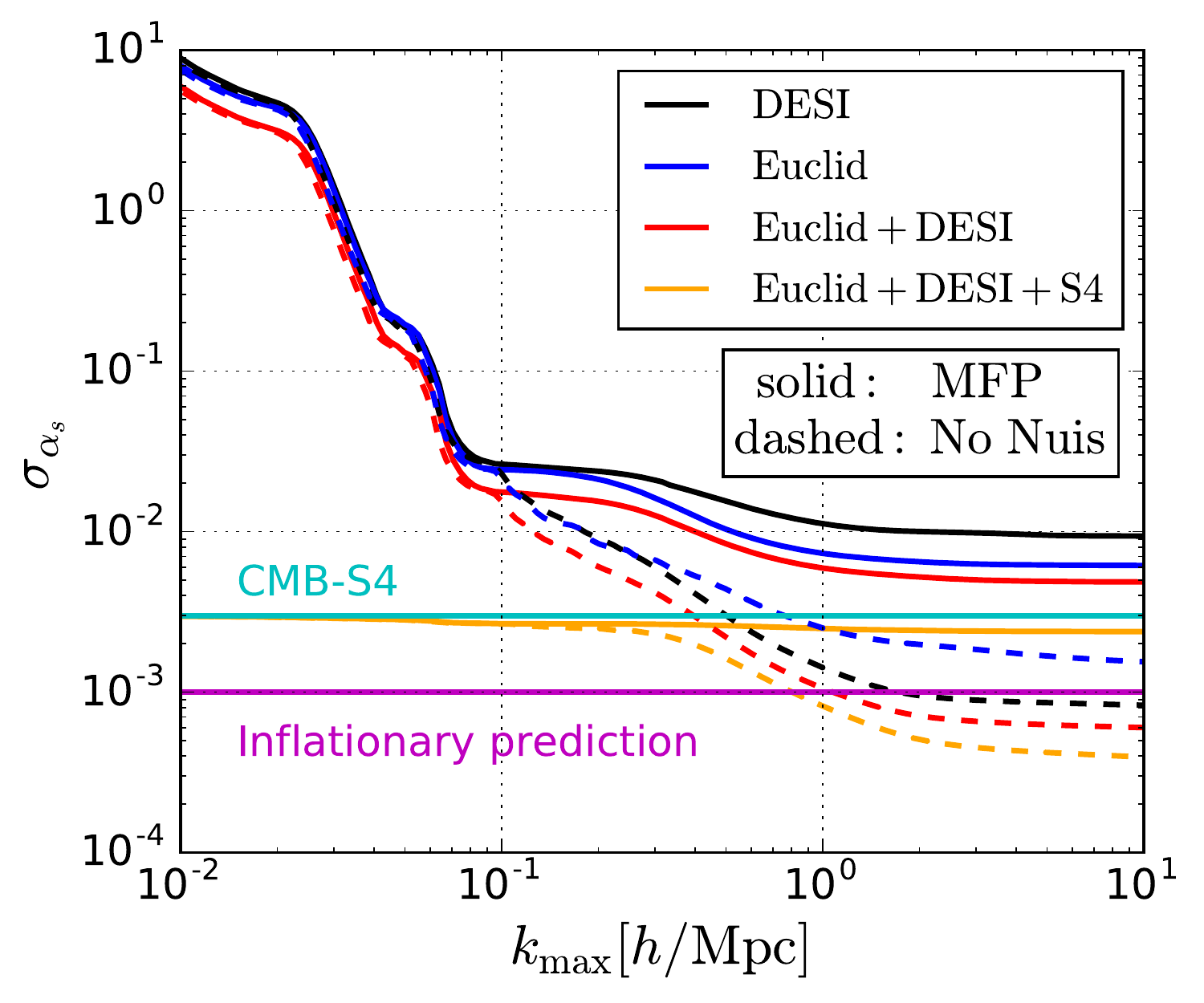}
\caption{Marginalized constraints on $\alpha_s$  when combining information from different surveys (DESI-like, Euclid-like, CMB-S4, and also in combination). Solid curves include the MFP description of the systematic errors in galaxy surveys (see Eq.~(\ref{eq:MFP}), while the dashed curves do not. Results using the Mead parameterization are similar to the $\nn$ (No Nuis) case, and so we we omit them here for clarity.}\label{fig:combine}
\end{figure*}

\begin{figure}[t]
\includegraphics[width=\linewidth]{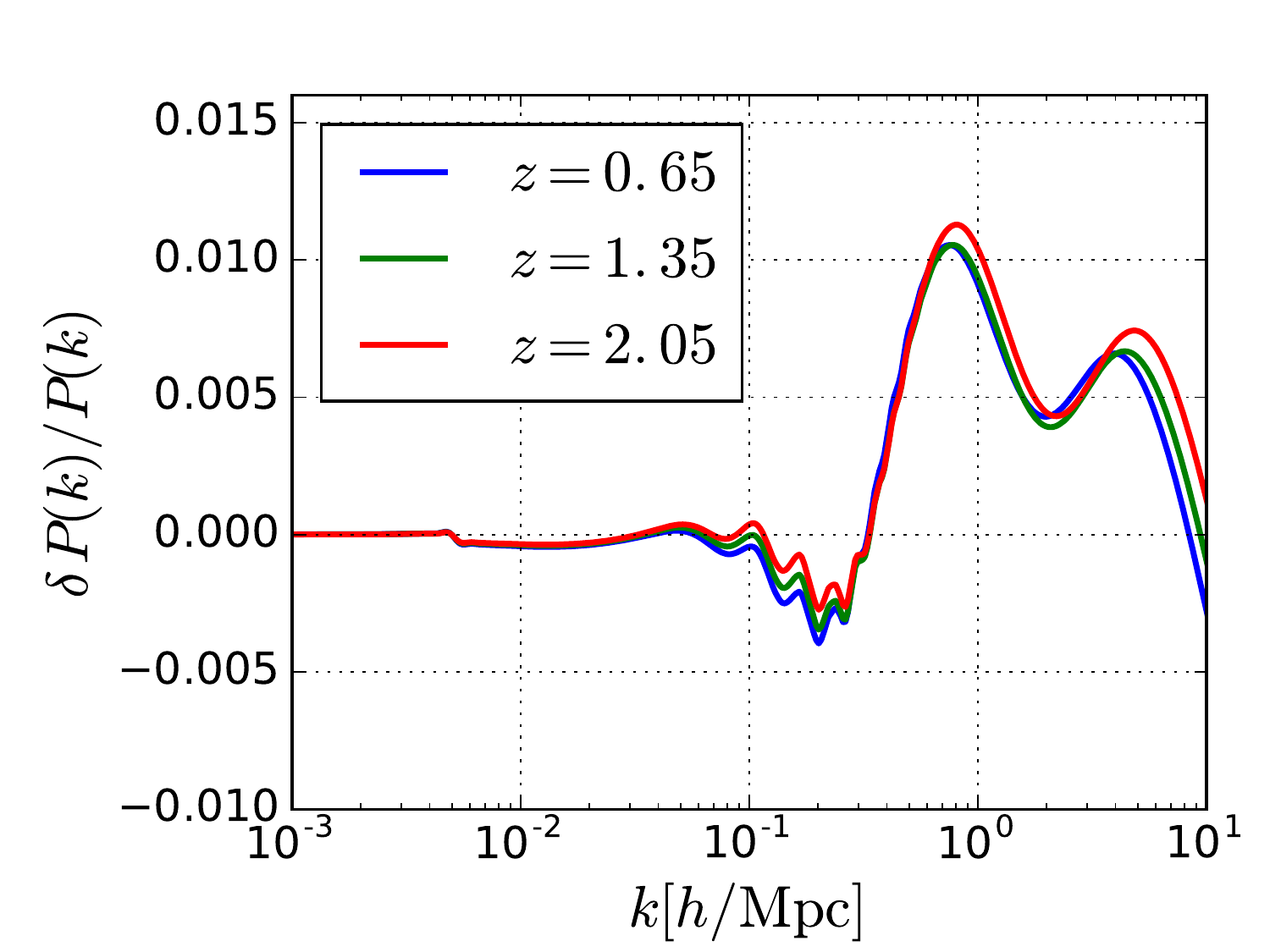}
\caption{Relative difference between the nonlinear predictions from two popular fits: that of~\citet{takahashi2012revising} and of~\citet{mead2015accurate} scaled so as to correspond to about 1\% maximum difference at small scales. The quantity shown, $\delta P/P = 0.2(P_\mathrm{taka}-P_\mathrm{mead})/P_\mathrm{mead}$, is the fiducial model for the small-scale systematics that we employ in subsequent plots to gauge the protection offered by our systematics parametrizations.
}\label{fig:deltap_p}
\end{figure}

\subsection{Galaxy Clustering}

To see how information from LSS data at small scales impacts constraints on the first spectral running, we forecast the marginalized $1\-\sigma$ constraints on $\alpha_s$ as a function of $\kmax$. For the moment, we hold the second running $\beta_s$ fixed at $0$; we will let $\beta_s$ vary further below, in Sec.~\ref{sec:beta}. 

Fig.~\ref{fig:error_Euclid} shows the increase in constraining power when we include clustering information at small scales, comparing the performance of the {$\nn$} (blue), Mead (red), and MFP models (black) for nonlinear effects. We also show constraints for the $\nn$ and MFP cases without the trispectrum contribution to the covariance (dashed), to demonstrate that its contribution to the error budget is minor (see Appendix~\ref{app:1000nz} for a case where shot noise is suppressed and the trispectrum dominates the  error budget).

In the {$\nn$} case, that is, the forecast for constraints if no parameters need to be introduced to model nonlinear effects, we find a large gain in constraining power for high $\kmax$. This gain remains whether or not we include the trispectrum contributions to the power spectrum's covariance at small scales (solid and dashed curves, respectively). 
However, the overall gain with increasing $\kmax$, even in this no-systematics case, is not as significant as might be expected  based on the behavior at linear scales, because the slope of the $\siga$ vs $\kmax$ curve changes at scales where nonlinearities become important, $\kmax\simeq 0.1\hmpcinv$. This flattening in $\siga(\kmax)$ implies that, even in the optimistic no-systematics scenario and pushing out to $\kmax=10\hmpcinv$, the Euclid  constraint on the running would only be comparable to the expected inflationary signal, $\siga\simeq 10^{-3}$ and so be insufficient for a statistically significant detection of $\alpha$ of that size. 

The red and black curves in Fig.~\ref{fig:error_Euclid} show how these constraints respond to the addition of nuisance parameters intended to capture nonlinear effects, corresponding to the Mead and MFP models, respectively. The Mead model, which introduces only two new parameters, produces results similar to the $\nn$ case. In contrast, constraints become considerably weaker (e.g.\ by a factor of $\sim$5 at $\kmax=10\hmpcinv$) for the MFP model, which captures nonlinear effects via an agnostic, piecewise-in-$k$ $\Mkz$ with many free parameters (up to 279 for the highest $\kmax$). Thus, in this more conservative treatment of small-scale systematics, the gains from including high-$k$ modes are rendered modest at best, particularly for $\kmax\gtrsim 1\hmpcinv$. We will show below in Sec.~\ref{sec:sys}, however, that the MFP parametrization does protect the constraints against the systematic biases due to modeling uncertainties in the high-$k$ power spectrum. 

Clearly, in the comparison of forecasted constraints, the more gentle treatment of systematics (with fewer free parameters) in the Mead model produces more favorable results than the more agnostic MFP case. However, this comparison of statistics-only errors alone is not enough to answer the question of whether a given treatment of systematics is sufficient for an analysis. Rather, modeling choices must be made by balancing the consideration of expected constraining power with the need for nuisance parameters to protect against biases to the best-fit cosmological parameters. Accordingly, we compare our three $\Mkz$ treatments by studying their relative ability to protect against biases in Sec.~\ref{sec:sys}.

\subsection{Galaxy clustering plus CMB}

The large lever arm provided by the combination of CMB and LSS allows for much tighter constraints on the running than using LSS data alone. We illustrate this in Fig.~\ref{fig:combine}, which gives the marginalized $1$-$\sigma$ constraints on $\alpha_s$ for different $\kmax$ when combining LSS information from a Euclid-like and/or a DESI-like survey with that from a CMB-S4 experiment. We now show only forecasts which include the trispectrum contribution to the covariance, and use the comparison between the solid and dashed curves to compare the performance of the {$\nn$} and MFP models, respectively. For clarity, we do not diplay the curves for the Mead model, which are similar to those for the {$\nn$} case.

The curves for LSS data alone show results similar to those in Fig.~\ref{fig:error_Euclid}. We find that DESI and Euclid yield comparable errors in the running (with a $\sim$30\% smaller error for Euclid), with their combination giving a slight improvement over Euclid alone. 

As in the Euclid-only case, we see a 5--10$\times$ degradation in constraints if the MFP treatment of systematics is adopted compared to the $\nn$ case. We note that this degradation is greater for DESI (black) than for Euclid (blue).

Next we consider the effect of adding CMB-S4 information to the Euclid+DESI combination, which is shown in orange in Fig.~\ref{fig:combine}. When large and mildly nonlinear scales of the LSS ($\kmax\lesssim 0.5\hmpcinv$) are used, the CMB information dominates the (CMB+LSS) constraining power, and the combined error is essentially equivalent to that from CMB-S4 alone. At smaller scales, the LSS surveys help tighten constraints, but only in the $\nn$ case. In the MFP case, where many nuisance parameters are marginalized over,  LSS data adds little constraining power on $\alpha_s$ compared to  CMB-S4 data alone.

\section{Systematic biases in model parameters}\label{sec:sys}

The fact that there are significant modeling uncertainties associated with the theoretical prediction of galaxy clustering at small scales is our primary motivation for studying different choices of the $\Mkz$ function to describe nonlinear effects. 
Any analysis will have to make simplifying choices for how to model the physics of nonlinear structure growth, baryonic effects, and scale-dependent galaxy bias. To the extent that those choices provide an incomplete description of the underlying physics there will be inaccuracies in the theoretical prediction for the observed galaxy power spectrum. Here we examine how these  systematic errors---that is, residuals between the true and assumed power spectrum---impact parameter estimation for the spectral running. 

In order to characterize this, we represent a typical form for the residuals due to systematic errors by taking 
the \textit{difference} between two commonly used parameterizations of the matter power spectrum on small scales. Specifically, we subtract the nonlinear prescription by~\citet{takahashi2012revising} from that of~\citet{mead2015accurate}.\footnote{We take the default parameter values of $\Abary$ and $\etao$ corresponding to the DMONLY case in \texttt{HMcode} as of Feb. 2018, which includes the updates of Ref.~\cite{Mead:2016zqy}} 
The power spectra generated with these two codes differ by up to $\sim$5\%, roughly independent of redshift for the range considered. For the future surveys we consider, we optimistically assume that that theoretical advances will allow the small-scale power spectrum to be computed to an accuracy of about 1\%.
We therefore adopt a fifth of the Takahashi-Mead difference as our fiducial model for residual systematics, that is,
\begin{equation}
\delta P(k,\mu,z)\,=
0.2\left [P_{\rm Taka}(k, \mu, z)-P_{\rm Mead}(k, \mu, z)\right ],
\label{eq:delta_Pk}
\end{equation}
which we show in Fig.~\ref{fig:deltap_p} as a fraction of our fiducial power spectrum.

\begin{figure}[t]
\includegraphics[width=\linewidth]{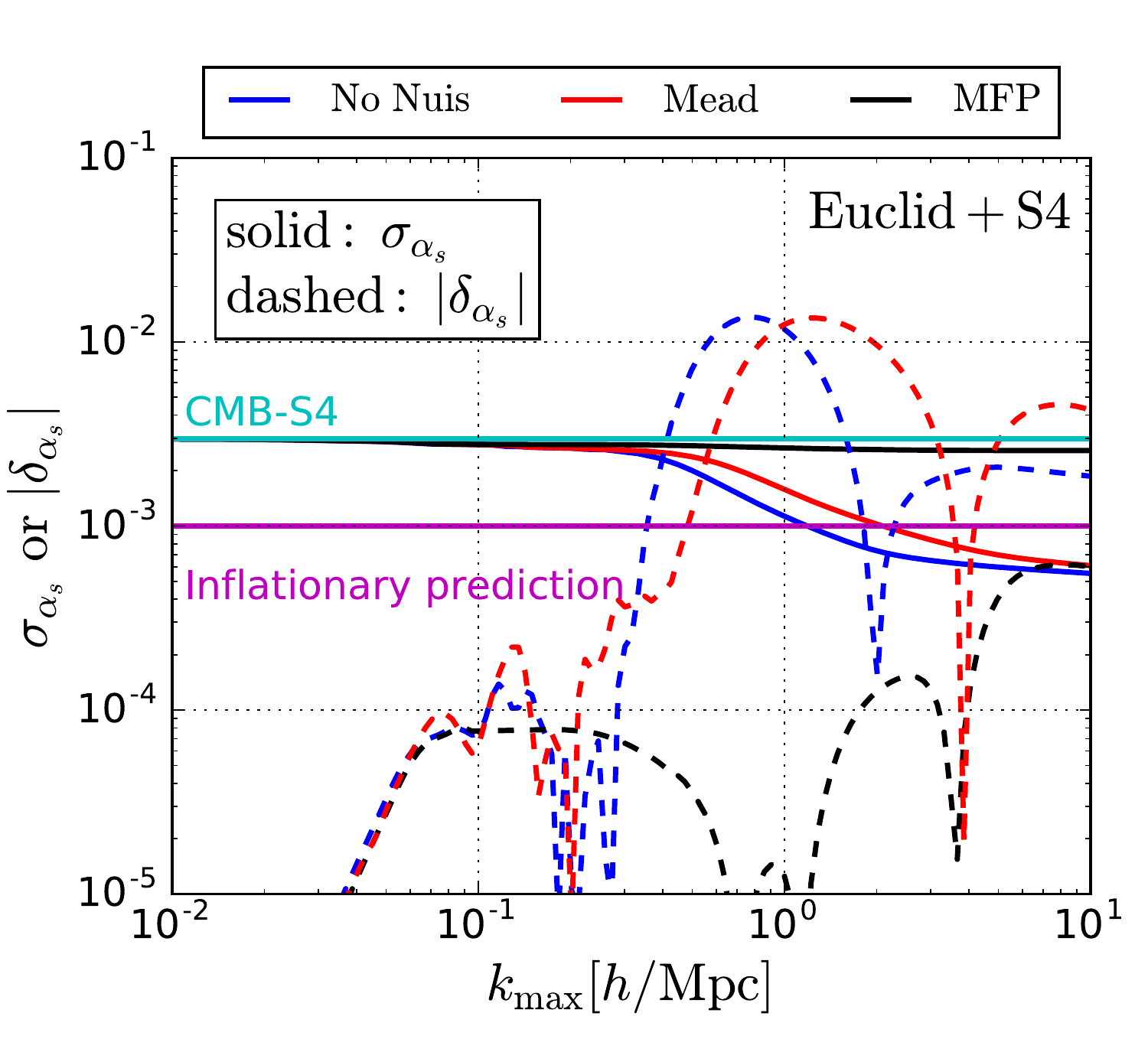}
\caption{1-$\sigma$ statistical errors (solid curves) and bias (dashed)  in the first spectral running, as a function of $\kmax$. We adopt the Euclid+CMB-S4 combination of surveys. The legend on top denotes three alternate assumptions about the systematic error modeling: none (blue), Mead (red), and MFP (black).}
\label{fig:bias_Euclid}
\end{figure}

We use the Fisher matrix formalism to predict the bias that the residuals in Eq.~(\ref{eq:delta_Pk}) will produce in cosmological parameters~\citep{Knox:1998fp,Huterer:2001yu}.  
In the limit where changes to best-fit parameters can be expanded linearly in small changes to the observations, the bias in parameter $p_i$ can be written as
\be \label{eq:bias}
\delta p_i \approx \sum_j(F^{-1})_{ij} G_j,
\ee
where 
\begin{align}
G_j = \sum_{z,\mu, {k_{\alpha}},k_{\beta}} d\mu \frac{\partial P(k_{\alpha},\mu,z)}{\partial p_j} \left[\rm{Cov}^{-1}\right]_{k_\alpha,k_\beta}\nonumber {\delta P(k_{\beta},\mu,z)},
\end{align}
and $\rm{Cov}$ is the same covariance matrix defined in Eq.~(\ref{eq:Cov}).  This formula is only accurate when the biases are small compared to the forecasted errors---that is, $|\delta p_i| \ll \sigma_{p_i} = \sqrt{(F^{-1})_{ii}}$---so we use it to determine the approximate threshold at which the bias on $p_i$ becomes unacceptably large. 

We plot both the bias $|\delta_{\alpha_s}|$ (dashed) and marginalized uncertainty $\sigma_{\alpha_s}$ (solid) for Euclid+CMB-S4 constraints on $\alpha_s$ in Fig.~\ref{fig:bias_Euclid}. The value of $\kmax$ where the bias and uncertainty become comparable tells us roughly the smallest scales that can be in included in an analysis without the systematic effects in $\delta P$ adversely biasing the results for $\alpha$. We see that though the MFP nuisance parameter prescription (black) has weaker constraints than the $\nn$ and Mead cases, it also is significantly better at protecting against bias. That is to say, on all $\kmax$ scales we examined, the bias in $\alpha$ for the MFP case is well below its statistical uncertainty. In contrast, the $\nn$ and Mead prescriptions have  $\delta_{\alpha_s}\approx\sigma_{\alpha_s}$ at $\kmax \approx 0.4\hmpcinv$ and $\kmax \approx 0.6\hmpcinv$ respectively. Comparing the value for $\sigma_{\alpha_s}$ at these $\kmax$ values, we see that if we restrict ourselves to scales with $\delta_{\alpha_s}<\sigma_{\alpha_s}$, the improvement from adding high-$k$ LSS data is marginal for all three $\Mkz$ treatments. 

To confirm that these results are robust against changes to the shape of our residual function $\delta P(k,z)$, we compared the same bias projections for a variety of other $\delta P_{i,j}(k,z) \propto P_i(k,z) - P_j(k,z)$, where $i,j \in \{$Mead~\citep{mead2015accurate}, Takahashi~\citep{takahashi2012revising}, Bird~\citep{Bird:2011rb}, Peacock\footnote{http://www.roe.ac.uk/~jap/haloes}, Halomodel~\citep{Peacock:2000qk}$\}$ runs over a subset of possible prescriptions for the nonlinear matter power spectrum in \texttt{CAMB}. 
We normalized these so that the relative difference $\delta P_{i,j}(k,z)/P_\textrm{Mead}$ had the same RMS as our fiducial case\footnote{For $k>0.005\hmpcinv$, corresponding to the minimum $k$ for which \texttt{CAMB} calculates nonlinear modifications to the power spectrum.} (see Appendix~\ref{app:dPk}). Thus the fiducial $\delta P(k,\mu,z)$ given in Eq.~\ref{eq:delta_Pk} and the magnitude of resulting biases derived therefrom should be fairly representative of possible errors in modeling $P(k,z)$, while also aligning with the oft-quoted baseline assumption that uncertainties have to be controlled to 1\% or better in order to not degrade the accuracy of future cosmological measurements of dark energy (e.g.~\cite{Huterer:2004tr}).

\section{Constraining the second running:  $\Lambda$CDM $+\alpha_s+\beta_s$}\label{sec:beta}

\begin{figure*}
\centering
\includegraphics[width=0.47\textwidth]{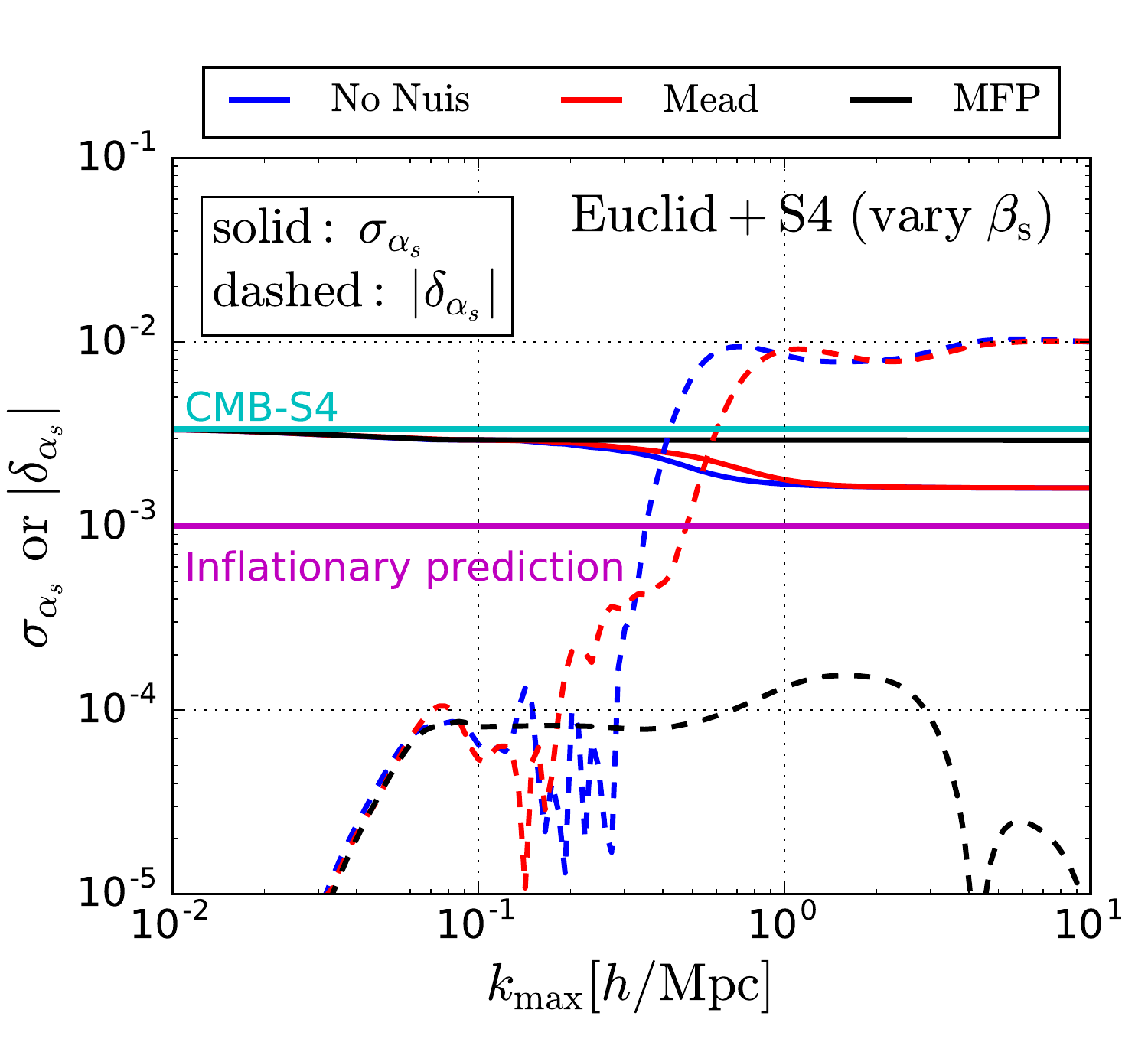}
\includegraphics[width=0.47\textwidth]{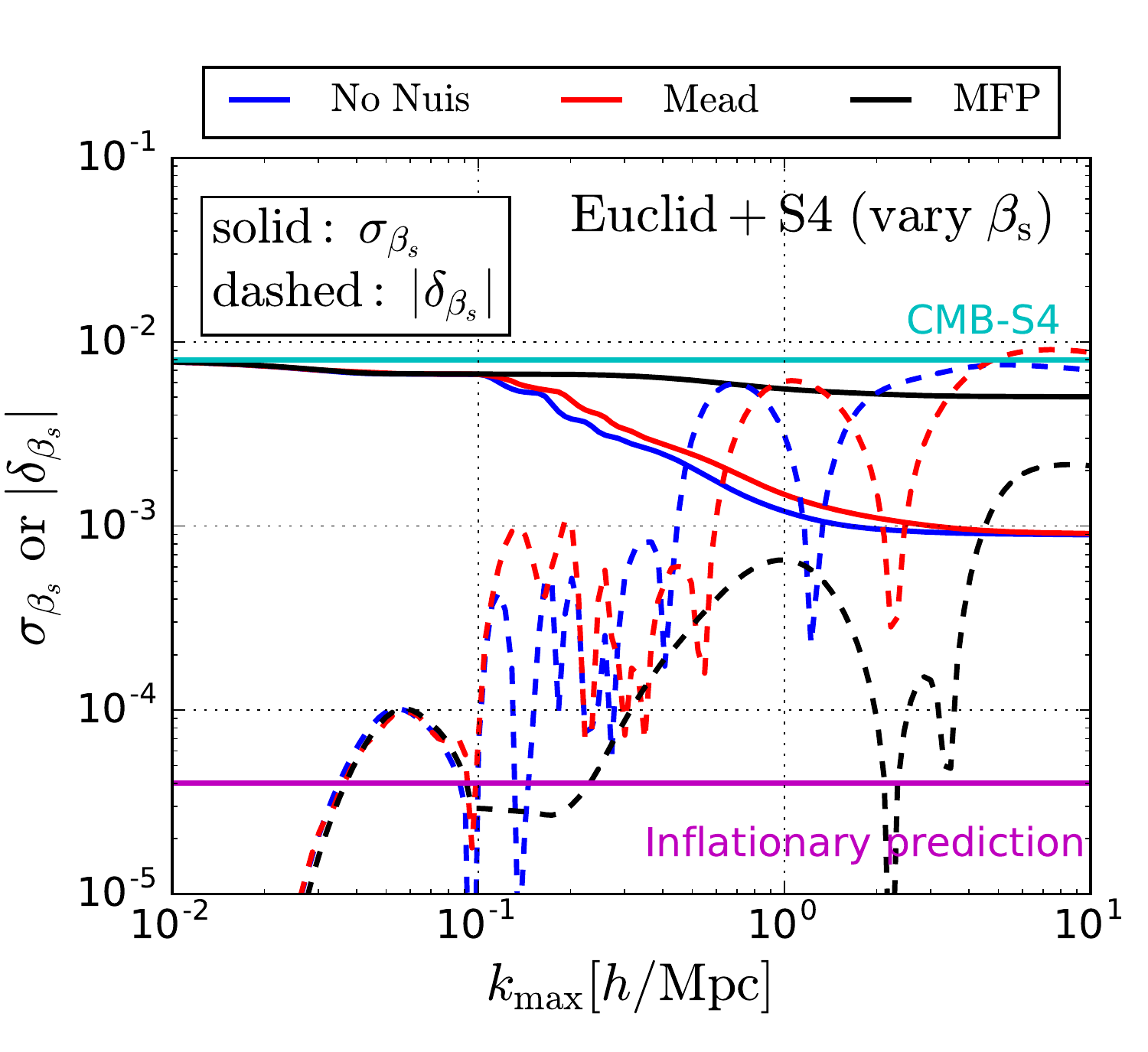}
\caption{Similar to Fig.~\ref{fig:bias_Euclid}, except now $\beta_s$ is allowed to vary. The left panel shows the 1-$\sigma$ error and parameter bias in $\alpha_s$ as a function of $\kmax$, while the right panel shows the same for $\beta_s$. The curves have the same meaning as  in Fig.~\ref{fig:bias_Euclid}.}
\label{fig:vary_bs}
\end{figure*}

We now expand the cosmological parameter space to include the second running $\beta_s$---that is, we extend our expansion of the spectral index to second order in $\ln k$. 
This is a parameter for which constraints from LSS data have the potential to be particularly interesting. Recent Planck results have suggested a positive second running $\beta_s$ at nearly $2\-\sigma$ confidence which, if it persists, will help to discriminate between inflationary models~\citep{Escudero:2015wba, Cabass:2016ldu}. Additionally, as mentioned in Sec.~\ref{sec:intro}, the current best-fit of $\beta_s = 0.025\pm0.013$ has important implications for physics of the late universe, as it makes primordial black holes a viable dark matter candidate (albeit with the requirement of a negative third-order running to avoid overproduction~\citep{Carr:2016drx}). 

The left panel of Fig.~\ref{fig:vary_bs} shows that, when $\beta_s$ is allowed to vary, combined constraints from CMB-S4 and LSS are no longer able to reach the inflationary prediction for $\alpha_s$ at any $\kmax < 10\hmpcinv$, even when the non-linear $P(k,z)$ is modeled perfectly and with no nuisance parameters (solid blue curve). On the other hand, the right panel of Fig.~\ref{fig:vary_bs} shows that $\beta_s$ itself benefits greatly from the addition of the LSS data.
While CMB-S4 is expected to improve constraints on $\beta_s$ by a factor of $\sim$4 over current levels, our results indicate that LSS data in the nonlinear regime from Euclid or DESI has the potential to improve this significantly up to $\kmax \sim 2\hmpcinv$, at which point shot noise limits the information that can be gained. 

We next consider the systematic biases in $\beta_s$ using the same prescription as in Sec.~\ref{sec:sys}. Using our fiducial model for power spectrum residuals due to unaccounted-for systematics [Eq.~(\ref{eq:delta_Pk})], Fig.~\ref{fig:vary_bs} shows that, without introducing undue bias, adding data from a Euclid-like survey can improve constraints on $\beta_s$ by a factor of 3--4 compared to the CMB-S4 only case.\footnote{This was the one case where our fiducial $\delta P(k,\mu,z)$ differed somewhat in its bias forecast from the ensemble of other $\delta P(k,\mu,z)$ tested, with $\delta_\beta/\sigma_\beta=1$ occurring at $\kmax\approx0.5$ and $0.7\hmpcinv$ for the $\nn$ and Mead models, respectively ($\sim4\times$ improvement in $\sigb$), compared to $\kmax\approx0.2$ and $0.4\hmpcinv$ for the typical $\dPk$ ($\sim3\times$ improvement in $\sigb$). The results are still qualitatively similar, however.} While still an order of magnitude too large to reach $\beta_s$ predicted by standard single-field slow-roll inflation, this level of precision is in the regime necessary to test for  models relevant for PBH formation~\citep{Carr:2016drx, munoz2017towards, Kohri:2018qtx}.

\section{Conclusions}\label{sec:conclusions}

In this work, we have investigated how small-scale information from large-scale structure surveys can improve constraints on the first [$\alpha_s$] and second [$\beta_s$] runnings of the scalar spectral index [$n_s$]. Previous analyses have been limited to the linear regime where the matter power spectrum is accurately described by theory, but the possibility of extending analyses to nonlinear regimes in the future is attractive. This is for two reasons:
First, there are many more modes at small scales and hence statistical errors from cosmic variance are greatly reduced.
Second, accessing high $k$ values provides a longer lever arm when combined with CMB constraints, which increases the sensitivity to variations in  the spectral index and its runnings. 

Attempts to include small-scale information are limited by challenges associated with theoretical modeling of the nonlinear power spectrum.  Nonlinear clustering of dark matter, baryonic effects, and scale-dependent galaxy bias all contribute to modeling uncertainties on small scales. Therefore, it is critical to not only calibrate models for these effects as accurately as possible, but also to carefully characterize how analyses' cosmological results are affected by residual errors in predictions for small-scale power.

Motivated by these considerations, we  compare forecasted constraints for spectral runnings from a few different parameterizations intended to capture the effects of systematics in the nonlinear regime. Specifically, we study cases where small scales are modeled using the  fiducial halo model code ($\nn$ case), the parameterization from~\citet{Mead:2016zqy} which introduces two nuisance parameters (Mead case), and an agnostic treatment adapted from~\citet{bielefeld2015cosmological} with up to a few hundred parameters, depending on $\kmax$ (Many Free Parameters, or MFP case). 

We first study the forecasts for statistical errors on the first spectral running $\alpha_s$ for future LSS surveys like Euclid and DESI alone, as well as in combination with CMB-S4. 
We find that in the $\nn$ and Mead cases, the constraints from  large-scale structure surveys tighten substantially as $\kmax$ is raised to include nonlinear scales. The MFP case also shows improvement, but with a flatter dependence on $\kmax$ and weaker constraints overall.  It is also at nonlinear scales where constraints using LSS and CMB data begin to improve $\alpha_s$ constraints compared to CMB-S4 data alone.  The tightest constraints come from the Euclid+DESI+CMB-S4 combination, for which our $\nn$ forecasts for statistical errors reach a value about a third of the $\alpha_s$ predicted by single-field slow-roll inflation at $\kmax\gtrsim 3\hmpcinv$. This could be precise enough to achieve a $\sim 3\sigma$ detection. 
These results become less promising, however, when we investigate the extent to which mismodeling of the nonlinear power spectrum biases cosmological parameter estimation. 
Using the difference between two commonly used nonlinear prescriptions as an example of expected modeling uncertainties, we determine the highest $\kmax$ we can use in an analysis before the resulting systematic bias in $\alpha_s$ becomes comparable to its statistical errors. 

We find that for 1\% errors in the power spectrum, in the $\nn$ case both $\alpha_s$ and $\beta_s$ remain unbiased (i.e.\ bias is smaller than the 1$\sigma$ statistical error) up to $\kmax\approx 0.3-0.4\hmpcinv$. Including these smaller scales results in significant improvements in $\sigb$, but only marginal improvements in $\siga$. 
Adopting the Mead parametrization of the systematics leads to very similar results 
indicating that the two free parameters from \citet{mead2015accurate}, motivated to account for baryonic feedback, are not sufficient to offer protection against the 1\%-level residual small-scale systematics in the power spectrum we might expect to encounter.  
In contrast, for the MFP parametrization $\alpha_s$ and $\beta_s$ are unbiased for all $\kmax$ studied, but the statistical error on the runnings in the CMB+LSS scenarios is only marginally better than that of CMB-S4 alone.

Our level of optimism regarding future measurements of the spectral runnings using LSS data is therefore mixed. The values of $\alpha_s$ and $\beta_s$ predicted by standard, single-field slow-roll models of inflation seem out of reach even when CMB-S4 information is combined with that of most powerful future LSS surveys once the small-scale systematics in the galaxy surveys are taken into account. On the other hand, larger values of spectral runnings predicted by other classes of inflationary models, as well as those motivated by other physics (e.g.\ primordial black holes) are within reach, and should be testable with the next generation of surveys.

\acknowledgments
XL is supported by China Scholarship Council, National Key R\&D Program of China No. 2017YFA0402600, the National Basic Science Program (Project 973) of China under Grant No. 2014CB845800, the National Natural Science Foundation of China under Grants Nos.11503001, 11633001 and 11373014, the Strategic Priority Research Program of the Chinese Academy of Sciences, Grant No. XDB23000000, the Interdiscipline Research Funds of Beijing Normal University, and the Opening Project of Key Laboratory of Computational Astrophysics, National Astronomical Observatories, Chinese Academy of Sciences.
DH and NW have been supported by DOE under contract DE-FG02-95ER40899. DH and SA have been supported by NASA under contract 14-ATP14-0005. JM has been supported by the Rackham Graduate School through a Predoctoral Fellowship.

\bibliography{ref}

\begin{thebibliography}{}
\expandafter\ifx\csname natexlab\endcsname\relax\def\natexlab#1{#1}\fi
\providecommand{\url}[1]{\href{#1}{#1}}

\bibitem[{Abazajian {et~al.}(2016)Abazajian, Adshead, Ahmed, Allen, Alonso,
  Arnold, Baccigalupi, Bartlett, Battaglia, Benson,
  {et~al.}}]{abazajian2016cmb}
Abazajian, K.~N., Adshead, P., Ahmed, Z., {et~al.} 2016, arXiv preprint
  arXiv:1610.02743

\bibitem[{Ade {et~al.}(2016{\natexlab{a}})}]{Ade:2015lrj}
Ade, P. A.~R., {et~al.} 2016{\natexlab{a}}, Astron. Astrophys., 594, A20

\bibitem[{Ade {et~al.}(2016{\natexlab{b}})}]{Ade:2015xua}
---. 2016{\natexlab{b}}, Astron. Astrophys., 594, A13

\bibitem[{Aghamousa {et~al.}(2016)Aghamousa, Aguilar, Ahlen, Alam, Allen,
  Prieto, Annis, Bailey, Balland, Ballester, {et~al.}}]{aghamousa2016desi}
Aghamousa, A., Aguilar, J., Ahlen, S., {et~al.} 2016, arXiv preprint
  arXiv:1611.00036

\bibitem[{Aghanim {et~al.}(2016)}]{Aghanim:2016yuo}
Aghanim, N., {et~al.} 2016, Astron. Astrophys., 596, A107

\bibitem[{Alam {et~al.}(2017)Alam, Ata, Bailey, Beutler, Bizyaev, Blazek,
  Bolton, Brownstein, Burden, Chuang, {et~al.}}]{alam2017clustering}
Alam, S., Ata, M., Bailey, S., {et~al.} 2017, Monthly Notices of the Royal
  Astronomical Society, stx721

\bibitem[{Albrecht \& Steinhardt(1982)}]{Albrecht:1982wi}
Albrecht, A., \& Steinhardt, P.~J. 1982, Phys. Rev. Lett., 48, 1220

\bibitem[{Albrecht {et~al.}(2009)}]{Albrecht:2009ct}
Albrecht, A., {et~al.} 2009, arXiv:0901.0721

\bibitem[{Amendola {et~al.}(2013)Amendola, Appleby, Bacon, Baker, Baldi,
  Bartolo, Blanchard, Bonvin, Borgani, Branchini,
  {et~al.}}]{amendola2013cosmology}
Amendola, L., Appleby, S., Bacon, D., {et~al.} 2013, Living Reviews in
  Relativity, 16, 6

\bibitem[{Bardeen {et~al.}(1983)Bardeen, Steinhardt, \&
  Turner}]{Bardeen:1983qw}
Bardeen, J.~M., Steinhardt, P.~J., \& Turner, M.~S. 1983, Phys. Rev., D28, 679

\bibitem[{Basse {et~al.}(2015)Basse, Hamann, Hannestad, \&
  Wong}]{Basse:2014qqa}
Basse, T., Hamann, J., Hannestad, S., \& Wong, Y. Y.~Y. 2015, JCAP, 1506, 042

\bibitem[{Bernardeau {et~al.}(2002)Bernardeau, Colombi, Gaztanaga, \&
  Scoccimarro}]{Bernardeau:2001qr}
Bernardeau, F., Colombi, S., Gaztanaga, E., \& Scoccimarro, R. 2002, Phys.
  Rept., 367, 1

\bibitem[{Bielefeld {et~al.}(2015)Bielefeld, Huterer, \&
  Linder}]{bielefeld2015cosmological}
Bielefeld, J., Huterer, D., \& Linder, E.~V. 2015, Journal of Cosmology and
  Astroparticle Physics, 2015, 023

\bibitem[{Bird {et~al.}(2012)Bird, Viel, \& Haehnelt}]{Bird:2011rb}
Bird, S., Viel, M., \& Haehnelt, M.~G. 2012, Mon. Not. Roy. Astron. Soc., 420,
  2551

\bibitem[{Blas {et~al.}(2011)Blas, Lesgourgues, \& Tram}]{blas2011cosmic}
Blas, D., Lesgourgues, J., \& Tram, T. 2011, Journal of Cosmology and
  Astroparticle Physics, 2011, 034

\bibitem[{Böhm {et~al.}(2016)Böhm, Schmittfull, \& Sherwin}]{Bohm:2016gzt}
Böhm, V., Schmittfull, M., \& Sherwin, B.~D. 2016, Phys. Rev., D94, 043519

\bibitem[{Cabass {et~al.}(2016)Cabass, Di~Valentino, Melchiorri, Pajer, \&
  Silk}]{Cabass:2016ldu}
Cabass, G., Di~Valentino, E., Melchiorri, A., Pajer, E., \& Silk, J. 2016,
  Phys. Rev., D94, 023523

\bibitem[{Carr {et~al.}(2016)Carr, Kuhnel, \& Sandstad}]{Carr:2016drx}
Carr, B., Kuhnel, F., \& Sandstad, M. 2016, Phys. Rev., D94, 083504

\bibitem[{Chisari {et~al.}(2018)Chisari, Richardson, Devriendt, Dubois,
  Schneider, Brun, Beckmann, Peirani, Slyz, \& Pichon}]{Chisari:2018prw}
Chisari, N.~E., Richardson, M. L.~A., Devriendt, J., {et~al.} 2018,
  arXiv:1801.08559

\bibitem[{Cooray \& Sheth(2002)}]{Cooray:2002dia}
Cooray, A., \& Sheth, R.~K. 2002, Phys. Rept., 372, 1

\bibitem[{Czerny {et~al.}(2014)Czerny, Kobayashi, \&
  Takahashi}]{Czerny:2014wua}
Czerny, M., Kobayashi, T., \& Takahashi, F. 2014, Phys. Lett., B735, 176

\bibitem[{Dodelson(2003)}]{Dodelson:2003ip}
Dodelson, S. 2003, AIP Conf. Proc., 689, 184, [,184(2003)]

\bibitem[{Drees \& Erfani(2012)}]{Drees:2011yz}
Drees, M., \& Erfani, E. 2012, JCAP, 1201, 035

\bibitem[{Easther \& Peiris(2006)}]{Easther:2006tv}
Easther, R., \& Peiris, H. 2006, JCAP, 0609, 010

\bibitem[{Eisenstein {et~al.}(1999)Eisenstein, Hu, \&
  Tegmark}]{eisenstein1999cosmic}
Eisenstein, D.~J., Hu, W., \& Tegmark, M. 1999, The Astrophysical Journal, 518,
  2

\bibitem[{Escudero {et~al.}(2016)Escudero, Ramírez, Boubekeur, Giusarma, \&
  Mena}]{Escudero:2015wba}
Escudero, M., Ramírez, H., Boubekeur, L., Giusarma, E., \& Mena, O. 2016,
  JCAP, 1602, 020

\bibitem[{{Evrard} {et~al.}(2008){Evrard}, {Bialek}, {Busha},
  {et~al.}}]{Evrard08}
{Evrard}, A.~E., {Bialek}, J., {Busha}, M., {et~al.} 2008, \apj, 672, 122

\bibitem[{Font-Ribera {et~al.}(2014)Font-Ribera, McDonald, Mostek, Reid, Seo,
  \& Slosar}]{font2014desi}
Font-Ribera, A., McDonald, P., Mostek, N., {et~al.} 2014, Journal of Cosmology
  and Astroparticle Physics, 2014, 023

\bibitem[{Garcia-Bellido \& Roest(2014)}]{GarciaBellido:2014gna}
Garcia-Bellido, J., \& Roest, D. 2014, Phys. Rev., D89, 103527

\bibitem[{Guth(1981)}]{Guth:1980zm}
Guth, A.~H. 1981, Phys. Rev., D23, 347

\bibitem[{Heitmann {et~al.}(2009)Heitmann, Higdon, White, Habib, Williams,
  Lawrence, \& Wagner}]{heitmann2009coyote}
Heitmann, K., Higdon, D., White, M., {et~al.} 2009, The Astrophysical Journal,
  705, 156

\bibitem[{Heitmann {et~al.}(2010)Heitmann, White, Wagner, Habib, \&
  Higdon}]{heitmann2010coyote}
Heitmann, K., White, M., Wagner, C., Habib, S., \& Higdon, D. 2010, The
  Astrophysical Journal, 715, 104

\bibitem[{Huang {et~al.}(2012)Huang, Verde, \& Vernizzi}]{Huang:2012mr}
Huang, Z., Verde, L., \& Vernizzi, F. 2012, JCAP, 1204, 005

\bibitem[{Huterer(2002)}]{Huterer:2001yu}
Huterer, D. 2002, Phys. Rev., D65, 063001

\bibitem[{Huterer \& Takada(2005)}]{Huterer:2004tr}
Huterer, D., \& Takada, M. 2005, Astropart. Phys., 23, 369

\bibitem[{Knox {et~al.}(1998)Knox, Scoccimarro, \& Dodelson}]{Knox:1998fp}
Knox, L., Scoccimarro, R., \& Dodelson, S. 1998, Phys. Rev. Lett., 81, 2004

\bibitem[{Kobayashi \& Takahashi(2011)}]{Kobayashi:2010pz}
Kobayashi, T., \& Takahashi, F. 2011, JCAP, 1101, 026

\bibitem[{Kohri \& Terada(2018)}]{Kohri:2018qtx}
Kohri, K., \& Terada, T. 2018, arXiv:1802.06785

\bibitem[{Kosowsky \& Turner(1995)}]{Kosowsky:1995aa}
Kosowsky, A., \& Turner, M.~S. 1995, Phys. Rev., D52, R1739

\bibitem[{Laureijs {et~al.}(2011)Laureijs, Amiaux, Arduini, Augueres,
  Brinchmann, Cole, Cropper, Dabin, Duvet, Ealet,
  {et~al.}}]{laureijs2011euclid}
Laureijs, R., Amiaux, J., Arduini, S., {et~al.} 2011, arXiv preprint
  arXiv:1110.3193

\bibitem[{Lawrence {et~al.}(2010)Lawrence, Heitmann, White, Higdon, Wagner,
  Habib, \& Williams}]{lawrence2010coyote}
Lawrence, E., Heitmann, K., White, M., {et~al.} 2010, The Astrophysical
  Journal, 713, 1322

\bibitem[{Lewis {et~al.}(2000)Lewis, Challinor, \& Lasenby}]{lewis538efficient}
Lewis, A., Challinor, A., \& Lasenby, A. 2000, Astrophys. J, 538, 473

\bibitem[{Lewis \& Pratten(2016)}]{Lewis:2016tuj}
Lewis, A., \& Pratten, G. 2016, JCAP, 1612, 003

\bibitem[{Linde(1982)}]{Linde:1981mu}
Linde, A.~D. 1982, Phys. Lett., B108, 389

\bibitem[{Mead {et~al.}(2016)Mead, Heymans, Lombriser, Peacock, Steele, \&
  Winther}]{Mead:2016zqy}
Mead, A., Heymans, C., Lombriser, L., {et~al.} 2016, Mon. Not. Roy. Astron.
  Soc., 459, 1468

\bibitem[{Mead {et~al.}(2015)Mead, Peacock, Heymans, Joudaki, \&
  Heavens}]{mead2015accurate}
Mead, A., Peacock, J., Heymans, C., Joudaki, S., \& Heavens, A. 2015, Monthly
  Notices of the Royal Astronomical Society, 454, 1958

\bibitem[{Mu{\~n}oz {et~al.}(2017)Mu{\~n}oz, Kovetz, Raccanelli, Kamionkowski,
  \& Silk}]{munoz2017towards}
Mu{\~n}oz, J.~B., Kovetz, E.~D., Raccanelli, A., Kamionkowski, M., \& Silk, J.
  2017, Journal of Cosmology and Astroparticle Physics, 2017, 032

\bibitem[{Peacock \& Smith(2000)}]{Peacock:2000qk}
Peacock, J.~A., \& Smith, R.~E. 2000, Mon. Not. Roy. Astron. Soc., 318, 1144

\bibitem[{Rudd {et~al.}(2008)Rudd, Zentner, \& Kravtsov}]{Rudd:2007zx}
Rudd, D.~H., Zentner, A.~R., \& Kravtsov, A.~V. 2008, Astrophys. J., 672, 19

\bibitem[{Samushia {et~al.}(2012)Samushia, Percival, \&
  Raccanelli}]{samushia2012interpreting}
Samushia, L., Percival, W.~J., \& Raccanelli, A. 2012, Monthly Notices of the
  Royal Astronomical Society, 420, 2102

\bibitem[{Scoccimarro {et~al.}(1999)Scoccimarro, Zaldarriaga, \&
  Hui}]{Scoccimarro:1999kp}
Scoccimarro, R., Zaldarriaga, M., \& Hui, L. 1999, Astrophys. J., 527, 1

\bibitem[{Seljak(1997)}]{seljak1997measuring}
Seljak, U. 1997, The Astrophysical Journal, 482, 6

\bibitem[{Seljak(2000)}]{Seljak:2000gq}
---. 2000, Mon. Not. Roy. Astron. Soc., 318, 203

\bibitem[{Seo \& Eisenstein(2007)}]{seo2007improved}
Seo, H.-J., \& Eisenstein, D.~J. 2007, The Astrophysical Journal, 665, 14

\bibitem[{Smith {et~al.}(2003)Smith, Peacock, Jenkins, White, Frenk, Pearce,
  Thomas, Efstathiou, \& Couchman}]{smith2003stable}
Smith, R.~E., Peacock, J.~A., Jenkins, A., {et~al.} 2003, Monthly Notices of
  the Royal Astronomical Society, 341, 1311

\bibitem[{Spergel {et~al.}(2015)}]{Spergel:2015sza}
Spergel, D., {et~al.} 2015, arXiv:1503.03757

\bibitem[{Takada \& Hu(2013)}]{Takada:2013bfn}
Takada, M., \& Hu, W. 2013, Phys. Rev., D87, 123504

\bibitem[{Takahashi {et~al.}(2012)Takahashi, Sato, Nishimichi, Taruya, \&
  Oguri}]{takahashi2012revising}
Takahashi, R., Sato, M., Nishimichi, T., Taruya, A., \& Oguri, M. 2012, The
  Astrophysical Journal, 761, 152

\bibitem[{{Tegmark} {et~al.}(1997){Tegmark}, {Taylor}, \&
  {Heavens}}]{tegmark1997fisher}
{Tegmark}, M., {Taylor}, A.~N., \& {Heavens}, A.~F. 1997, \apj, 480, 22

\bibitem[{Tegmark {et~al.}(2004)Tegmark, Strauss, Blanton, Abazajian, Dodelson,
  Sandvik, Wang, Weinberg, Zehavi, Bahcall, {et~al.}}]{tegmark2004cosmological}
Tegmark, M., Strauss, M.~A., Blanton, M.~R., {et~al.} 2004, Physical Review D,
  69, 103501

\bibitem[{van Daalen {et~al.}(2011)van Daalen, Schaye, Booth, \&
  Vecchia}]{vanDaalen:2011xb}
van Daalen, M.~P., Schaye, J., Booth, C.~M., \& Vecchia, C.~D. 2011, Mon. Not.
  Roy. Astron. Soc., 415, 3649

\bibitem[{Wu \& Huterer(2013)}]{wu2013impact}
Wu, H.-Y., \& Huterer, D. 2013, Monthly Notices of the Royal Astronomical
  Society, 434, 2556

\bibitem[{Zaldarriaga \& Seljak(1997)}]{zaldarriaga1997all}
Zaldarriaga, M., \& Seljak, U. 1997, Physical Review D, 55, 1830

\end{thebibliography}

\appendix

\begin{figure*}
\includegraphics[width=0.48\linewidth]{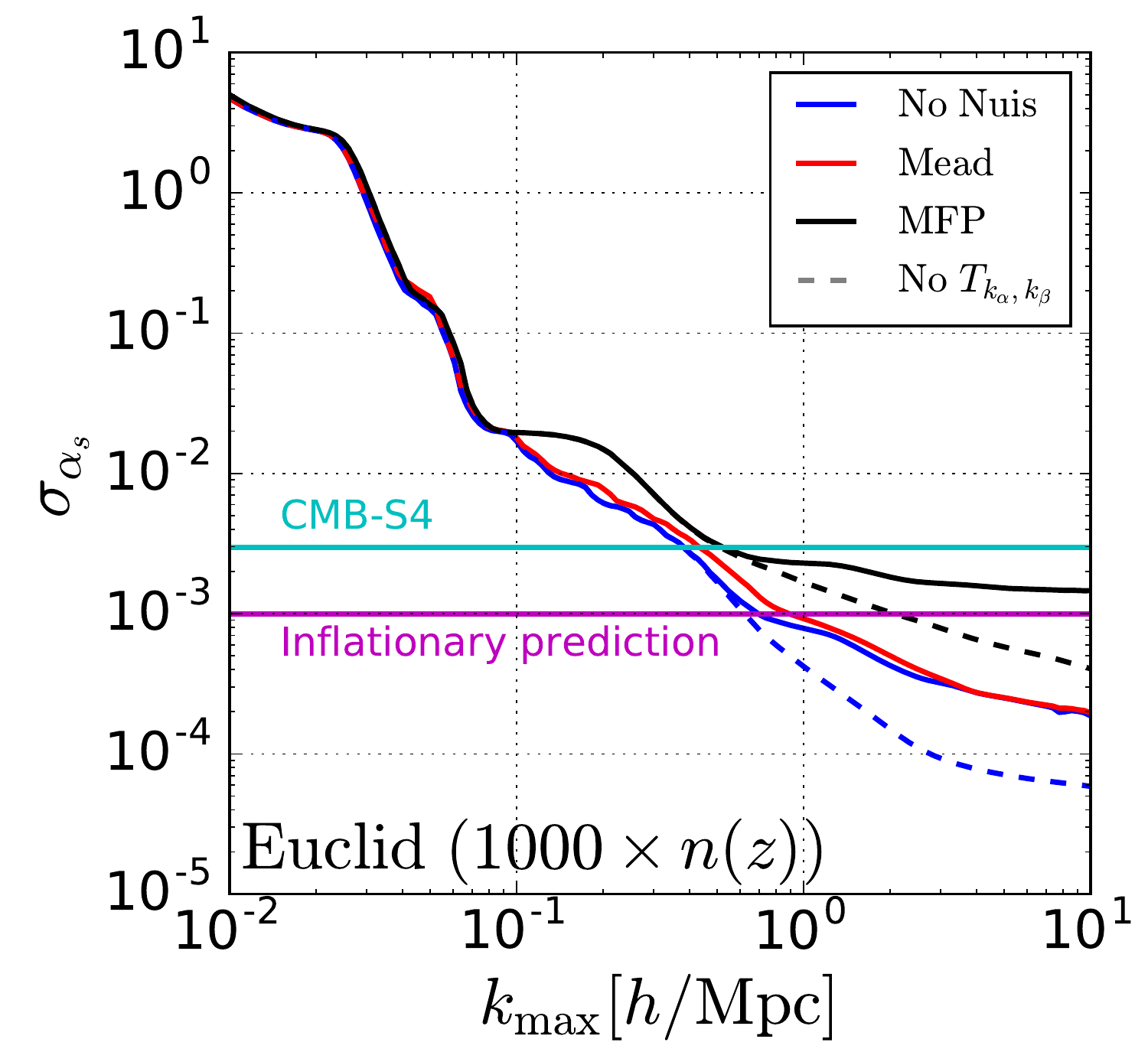}
\includegraphics[width=0.48\linewidth]{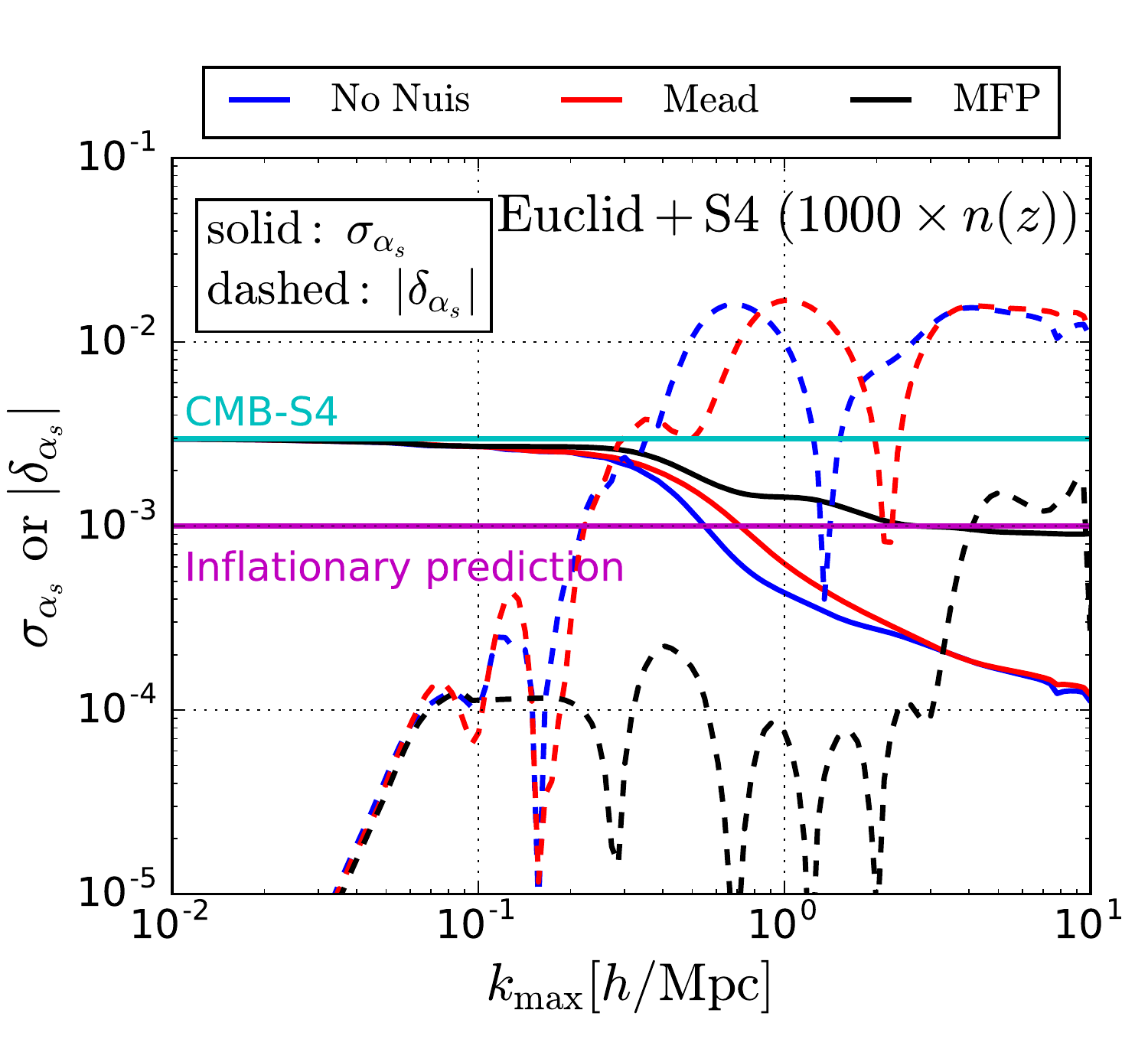}
\caption{Constraints on the spectral running $\alpha_s$ for a hypothetical survey with $n(z)\rightarrow 1000\times [n(z)_\mathrm{Euclid}]$ alone (left, compare to Fig.~\ref{fig:error_Euclid}) and with a CMB-S4 experiment (right, compare to Fig.~\ref{fig:bias_Euclid}). While the LSS constraints improve with the increased number density, the trispectrum still limits the information that can be gained from nonlinear scales of $k \gtrsim 0.6 \hmpcinv$ (\textit{left}, dashed vs. solid). If $P(k,z)$ is mismodeled, then only the MFP prescription (black) improves constraints over CMB-S4 before significantly biasing the results.}\label{fig:1000nP}
\end{figure*}

\section{Increasing the number density $n(z)$} \label{app:1000nz}

As shown in Figures~\ref{fig:error_Euclid} and \ref{fig:bias_Euclid}, a Euclid-like survey will be unable to constrain the spectral running to $\siga < 10^{-3}$, which would be necessary to be able to detect the value predicted by single-field, slow-roll inflation. 
To better understand the limiting factors of these forecasted constraints
we consider constraints for a survey similar to the Euclid-like one studied above, but with the number density increased dramatically to $n(z) \rightarrow 1000\times n(z)$. We show forecasts for its statistical errors and systematic biases in Fig.~\ref{fig:1000nP}. We find that for this high-source-density survey, the LSS information tightens constraints at lower $\kmax$, reaching $\siga\lesssim 10^{-3}$ at $\kmax \approx \{0.5, 0.7, 2\}\hmpcinv$ for the no nuisance, Mead, and MFP models, respectively. We also find that the increased density makes parameter estimation for $\alpha_s$ more sensitive to systematic biases: if $P(k)$ is mismodeled,  then only the MFP model improves constraints over CMB-S4 before introducing unacceptable levels of bias.

This hypothetical $1000\times n(z)$ survey is also useful to gauge the effect of the trispectrum-induced covariance on cosmological parameter constraints from modes in the strongly nonlinear regime. Unlike our main results in Figure~\ref{fig:error_Euclid}, where the covariance term was shot-noise dominated at small scales, the trispectrum term becomes important when the number density is very high. The result, as can be seen in Figure \ref{fig:1000nP}, is that there is little improvement in $\siga$---especially when combining with CMB-S4---from wavenumbers $k \gtrsim 2 \hmpcinv$. Note that we have not included the additional ``super-sample covariance" term~\citep{Takada:2013bfn} that could further degrade the contribution from modes in the nonlinear regime. 

Therefore we conclude that, once the realistic systematics are accounted for, even a Euclid-like survey with an artificially high number density of sources is unable to reach the precision required to detect the spectral runnings predicted by single-field, slow-roll inflationary models.

\section{Robustness of results to choice of $\dPk$} \label{app:dPk}
As noted in Sec.~\ref{sec:sys}, here we consider the robustness of our parameter bias results against changes to the shape of $\dPk$. We do this by  computing the differences between various prescriptions for the nonlinear power spectrum available in \texttt{CAMB}. 
Because we want to test sensitivity to the shape of $\dPk$, we normalize each curve so that its RMS over all $z$ and $0.005<k\leq 10\hmpcinv$ is equal to that of our fiducial ``takahashi-mead" $\dPk$. 

\begin{figure*}
\centering
\includegraphics[width=0.80\textwidth]{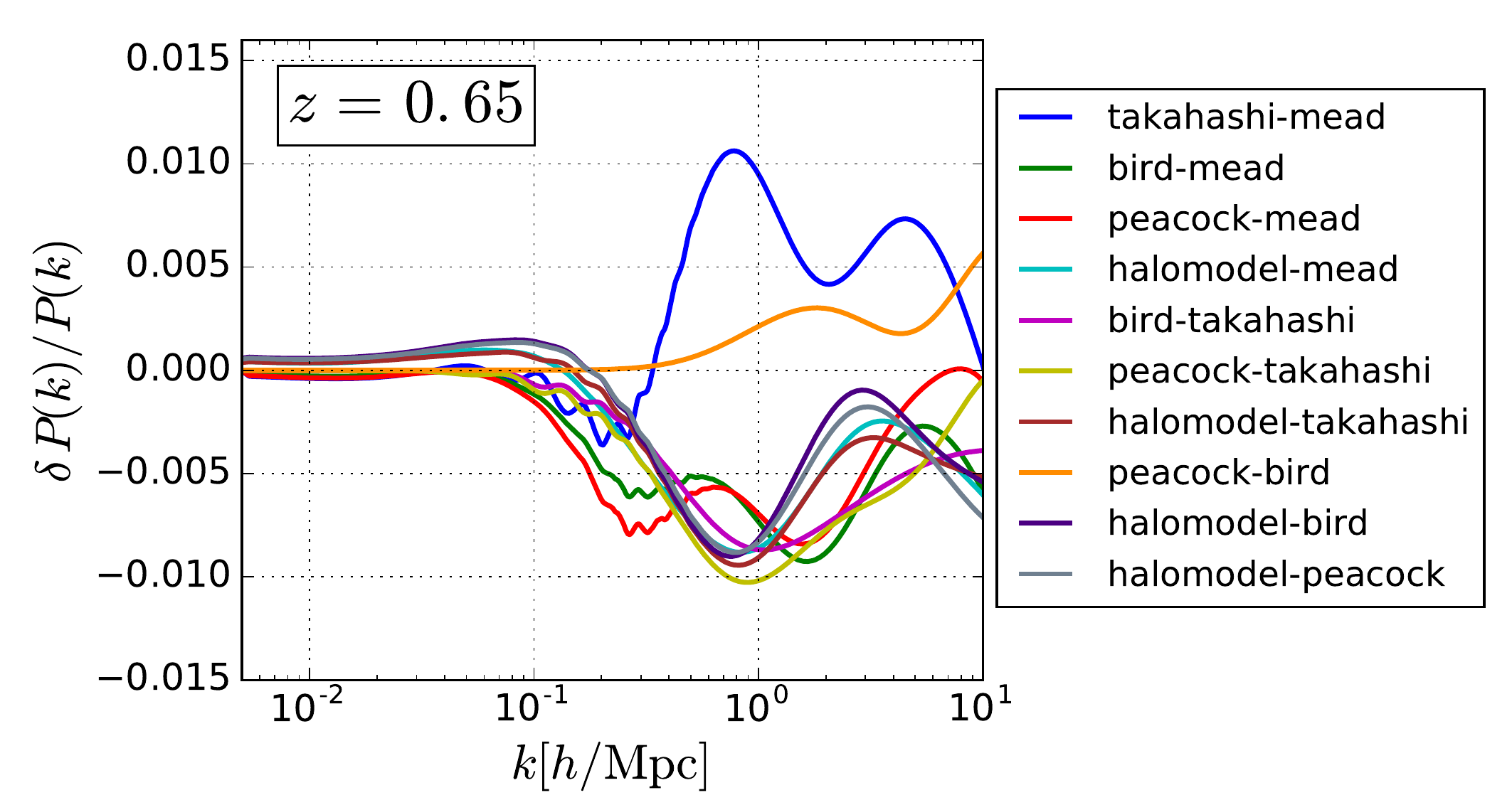}
\caption{Other systematic shifts in $P(k)$ tested to verify that the results of Sec.~\ref{sec:sys} are robust to choice of $\delta P(k)$. Note that because Eq.~(\ref{eq:bias}) is linear in $\delta P(k)$ and we are interested in where $|\delta|/\sigma=1$, the overall sign of $\delta P(k)$ is inconsequential. }
\label{fig:alldpk}
\end{figure*}

Fig.~\ref{fig:alldpk} shows the resulting ensemble of $\dPk$ considered, for the shallowest redshift bin, $z=0.65$. When looking at this Figure, there are a couple of things worth noting. First, because we are primarily interested in how constraints on the runnings become biased as we push to higher scales, i.e. $\kmax$ at which $|\delta_{p_i}|/\sigma_{p_i}=1$, the results are insensitive to the sign of $\dPk$. 
Second, the relatively small magnitude of the bird$-$peacock (orange)  curve is due to its large magnitude at higher redshifts compared to the other curves. Thus the low-$z$ range shown contributes less to its normalized RMS is less than it does for the other curves.

\begin{figure*}
\centering
\includegraphics[width=1.\textwidth]{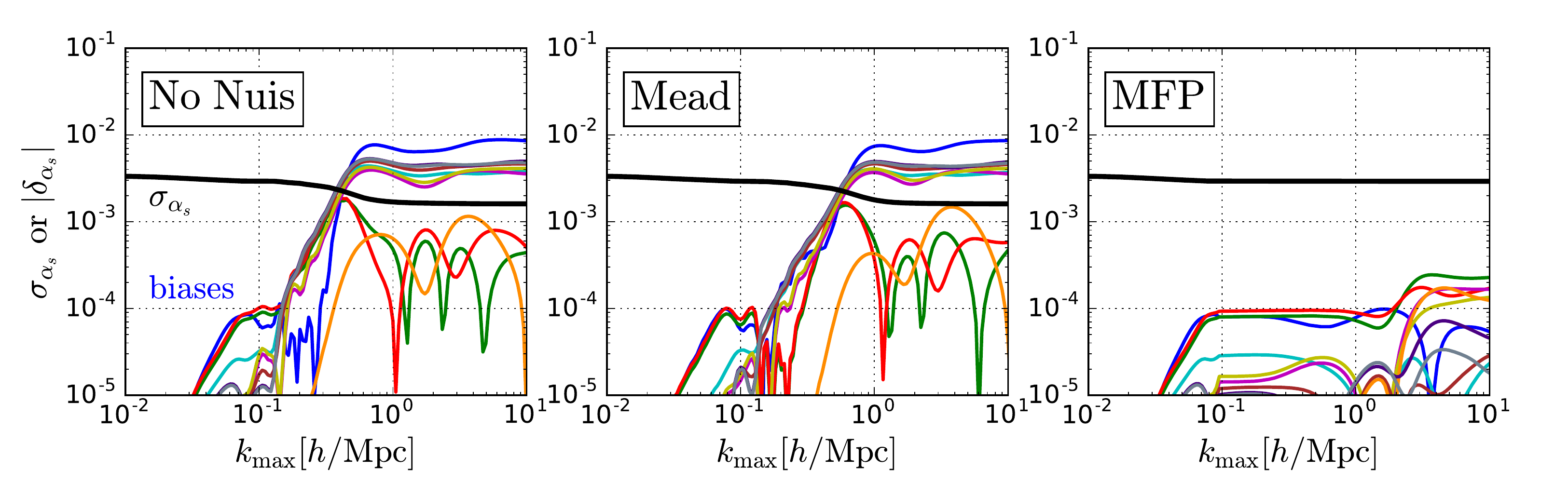}
\includegraphics[width=1.\textwidth]{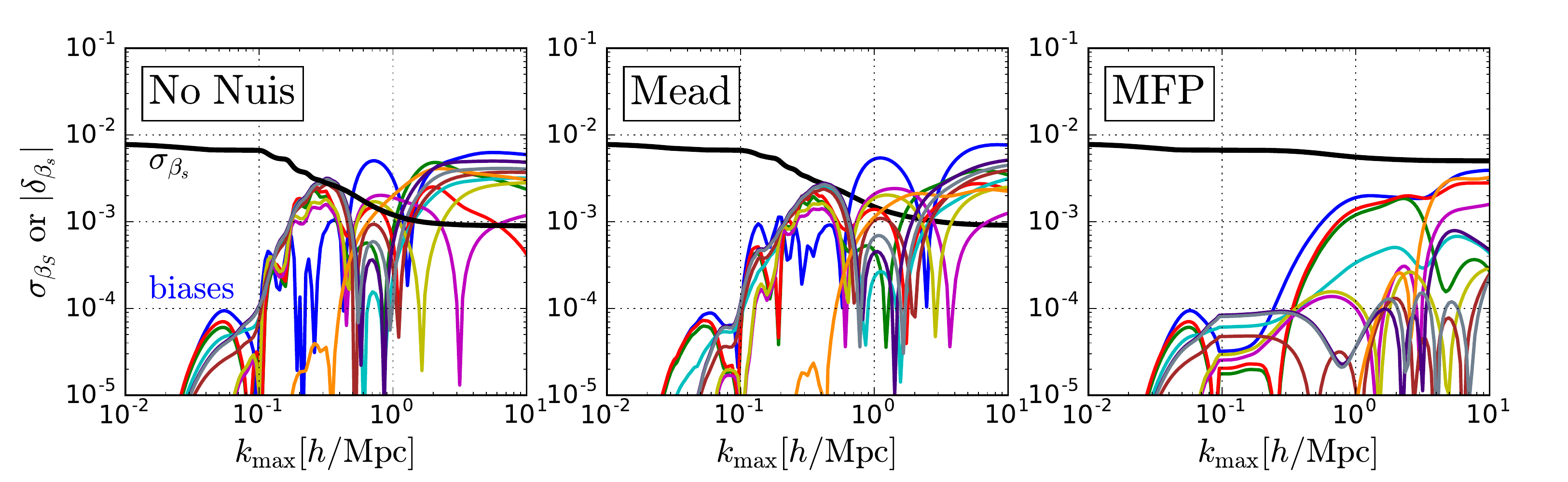}
\caption{Parameter bias from different $\delta P(k,z)$ for \lcdm$+\alpha_s+\beta_s$ using Euclid + CMB-S4 for $\alpha_s$ (top) and $\beta_s$ (bottom). The 1$\sigma$ uncertainty is in black and columns correspond to different nonlinear prescriptions from Sec.~\ref{sec:methods}.}
\label{fig:alldpkbias}
\end{figure*}

The parameter biases in $\alpha_s$ and $\beta_s$ resulting from these $\dPk$ curves are shown in Fig.~\ref{fig:alldpkbias} for the combined analysis of Euclid and CMB-S4.  These biases are analogous to those shown in Fig.~\ref{fig:vary_bs}. Though there is certainly variation in the shape of the curves, we see that the results for $\delta_{\alpha_s}(\kmax)$ and $\delta_{\beta_s}(\kmax)$  for our fiducial $\dPk$ (blue solid curves) are fairly typical.
Therefore, we conclude that our fiducial choice of the uncorrected bias in $P(k, z)$ at small scales, given in Eq.~(\ref{eq:delta_Pk}), is fairly typical of such choices. 

\end{document}